# Visualisation of multi-indication randomised control trial evidence to support decision-making in oncology: a case study on bevacizumab

Sumayya Anwer[1], Janharpreet Singh[2], Sylwia Bujkiewicz[2], Anne Thomas[3], Richard Adams[4,5], Elizabeth Smyth[6], Pedro Saramago[7], Stephen Palmer[7], Marta O Soares[7], Sofia Dias[1]

[1] Centre for Reviews and Dissemination, University of York, York, UK

[2] Biostatistics Research Group, Department of Population Health Sciences, University of Leicester, Leicester, UK

[3] Leicester Cancer Research Centre, University of Leicester, Leicester, UK

[4] Cardiff University, Cardiff, UK

[5] Velindre Cancer Centre, Cardiff, UK

[6] Oxford NIHR Biomedical Research Centre, Churchill Hospital, Oxford, OX3 7LE

[7] Centre for Health Economics, University of York, UK

## Abstract

**Background:** Evidence maps have been used in healthcare to understand existing evidence and to support decision-making. In oncology they have been used to summarise evidence within a disease area but have not been used to compare evidence across different diseases. As an increasing number of oncology drugs are licensed for multiple indications, visualising the accumulation of evidence across all indications can help inform policy-makers, support evidence synthesis approaches, or to guide expert elicitation on appropriate cross-indication assumptions.

**Methods:** The multi-indication oncology therapy bevacizumab was selected as a case-study. We used visualisation methods including timeline, ridgeline and split-violin plots to display evidence across seven licensed cancer types, focusing on the evolution of evidence on overall and progression-free survival over time as well as the quality of the evidence available.

**Results:** Evidence maps for bevacizumab allow for visualisation of patterns in study-level evidence, which can be updated as evidence accumulates over time. The developed tools display the observed data and synthesised evidence across- and within-indications.

**Limitations:** The effectiveness of the plots produced are limited by the lack of complete and consistent reporting of evidence in trial reports. Trade-offs were necessary when deciding the level of detail that could be shown while keeping the plots coherent.

**Conclusions:** Clear graphical representations of the evolution and accumulation of evidence can provide a better understanding of the entire evidence base which can inform judgements regarding the appropriate use of data within and across indications.

**Implications:** Improved visualisations of evidence can help the development of multi-indication evidence synthesis. The proposed evidence displays can lead to the efficient use of information for health technology assessment.

# 1 Introduction

Evidence maps are visual tools that can be used to systematically summarise existing evidence by displaying key characteristics such as study design, populations, interventions, comparators, and outcomes. These maps can provide a foundation for further, more focused, research synthesis by guiding stakeholders to high quality research, informing research priority setting and helping define the focus of evidence synthesis.[1] They can also be used to identify and highlight evidence gaps.[2] Within a healthcare context, evidence maps have been used, for example, to support decision-making in chemicals policy and risk management,[3] identify gaps in healthcare policy and governance in low and middle-income countries,[4] and understand the extent and distribution of evidence for interventions in youth mental health disorders.[5] Data visualisations may be static or interactive,[6] and can be used to support decision-makers and policymakers by highlighting relevant information such as public health indicators or social determinants of health.[6, 7] For instance, visualisations in the form of timelines have been used to represent trial design,[8] evidence availability over time,[9] and to depict patient care and diagnoses.[10]

An increasing number of oncology drugs are licensed for multiple indications, typically sequentially, so that a drug is licensed for a single indication initially and over time its license is extended to include additional indications. However, health technology assessment (HTA) bodies generally appraise drugs for one indication (the 'target' indication) at a time and ignore evidence from other indications across different disease areas.[11] The use of often immature evidence from only one, or very few indication-specific trials can result in uncertain treatment effect estimates. HTA-informed decisions about oncology treatments with evidence available from multiple indications (multi-indication) may be improved by making better use of evidence across- as well as within-indications.

Sharing of evidence from previously licensed indications in different disease areas can strengthen estimates for the target indication. Panoramic meta-analyses[15-17] can be used to pool treatment effects across as well as within indications, allowing for both between-and within-indication variation. However, judgements need to be made about the appropriateness of combining evidence across indications and it may be difficult to make these judgements without an effective way to visualise the existing evidence specific to the models we are interested in.

Attributes of multi-indication oncology evidence can introduce challenges in summarising and presenting evidence in ways that are useful in HTA. This includes the fact that two, related, time to event outcomes, progression-free survival (PFS) and overall survival (OS) are often of interest, with studies reporting one or both of these outcomes at multiple (interim as well as final), potentially different, time-points. Relative effectiveness estimates for the drug of interest compared to key comparators may be available, and relevant comparators will typically differ across indications.

Our paper develops novel visualisation tools to provide a comprehensive overview of the available evidence, with the aim of improving decision-making for a target indication. These visualisations can be used in a single as well as multiple indication context to show the evolution of evidence over time. We will explore displays of the aggregate level published evidence for each indication, as well as different ways to visualise the impact of making different assumptions and fitting across-indication synthesis models.

We will use the case study of bevacizumab (first licensed as Avastin®), to describe methods of visualising the available evidence for a technology across multiple indications. We selected bevacizumab, one of the first targeted multi-indication oncology therapies, as a case study as it has an extensive evidence base across multiple cancer indications over a period of more than twenty years. We aim to display how evidence accumulates over time and key evidence characteristics such as the magnitude and maturity of the estimated treatment effects when considered independently or after combining evidence across different indications. We will discuss how these displays can be used to help inform the judgements necessary to support the assumptions required for evidence synthesis models used to support HTA decisions and how they may be extended beyond this case-study.

## 2 Bevacizumab case-study: Establishing the evidence-base

Bevacizumab was the first available angiogenesis inhibitor therapy. It was licensed for the treatment of metastatic colorectal cancer in combination with chemotherapy in the US in 2004 and the European Union in 2005.[9] The National Institute for Health and Care Excellence (NICE) in the UK undertook the first UK HTA appraisal of bevacizumab for the treatment of metastatic colorectal cancer in 2007. Since its initial licensing, bevacizumab has received license extensions for a further six cancer types . We aimed to identify evidence on the relative treatment effects (RTE) comparing bevacizumab against alternative treatments in terms of OS and PFS. New and existing evidence displays are developed and adapted to illustrate the evolution of bevacizumab evidence over time, across its multiple licensed indications.

### 2.1 Study identification

To establish evidence on the indications for which bevacizumab is approved we used the summary of product characteristic (SmPC) for Avastin®, issued by the European Medicines Agency (EMA).[18] We identified seven licensed cancer types: breast cancer, cervical cancer, colorectal cancer, glioblastoma, non-small cell lung cancer (NSCLC), renal cell carcinoma, and ovarian cancer (which is the term used here to refer to ovarian, fallopian tube, and primary peritoneal cancers collectively).

Searching for evidence on these indications was conducted in two stages. In the first stage, we searched for all relevant comparative phase II or phase III randomised controlled trials (RCTs) of bevacizumab that were either included in NICE appraisals, the SmPC for Avastin® or in Cochrane reviews on the seven licensed cancer types. This was followed by a second search on the clinicaltrials.gov database[19] for phase III Avastin® trials that were either complete or had been terminated prior to completion. Any trials that had not been identified previously were included. We also identified and included studies from two relevant systematic reviews[9, 20] that were already known to us.

Inclusion Criteria

We included oncology studies in the metastatic/advanced setting where the treatment effect for bevacizumab could be isolated from any background chemotherapies or other targeted therapies administered during the trial.

Exclusion Criteria

We excluded studies in non-licensed indications and non-cancer therapeutic areas (e.g. macular degeneration). Studies where bevacizumab was administered in an adjuvant or neo-adjuvant setting were also excluded as the treatment effect of bevacizumab in these settings was expected to differ substantially from the advanced/metastatic setting.

Data extraction

For each selected trial, we retrieved all available publications (using clinicaltrials.gov records and checking citations for the main trial publication) and extracted data for all interim and final datapoints. Details of the data extraction process are included in Supplementary Section A-II.

The identification of studies is depicted in Figure 1. The final dataset consisted of 41 unique trials across the seven cancer types. A list of all relevant identified studies is included in Supplementary Section A, Table S1.

*Figure 1. PRISMA diagram for the study search process.*

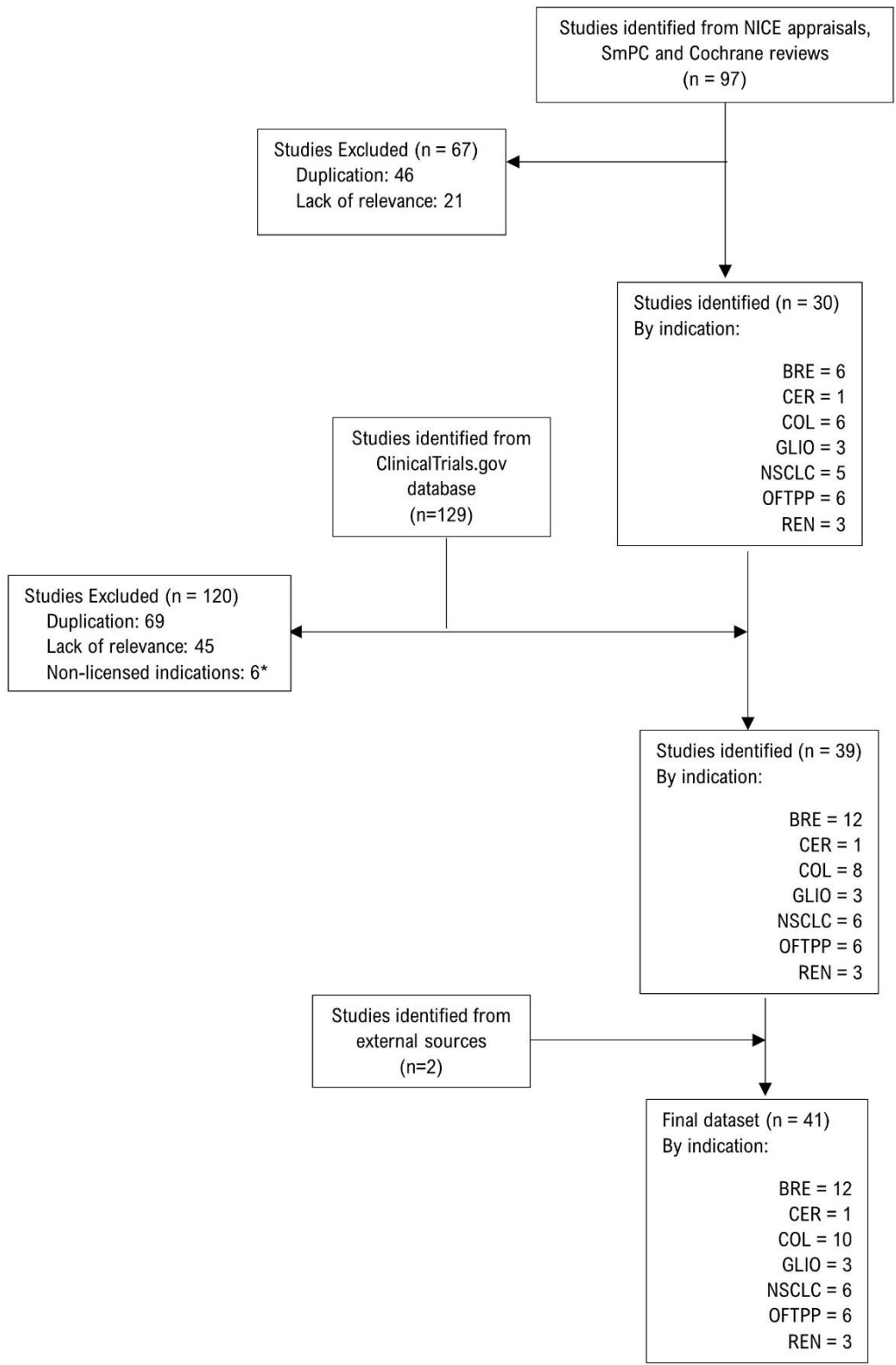

\* The non-licensed indications identified were lymphoma, gastrointestinal, urothelial, prostate and uterine cancer.

**Abbreviations:** BRE, breast cancer; CER, cervical cancer; COL, colorectal cancer; GLIO, glioblastoma; NSCLC, non-small cell lung cancer; OFTPP, ovarian, fallopian tube and primary peritoneal cancer; REN, renal cell carcinoma.

# 3 Oncology evidence data features to display in visualisations

Oncology data presents a set of features that are important to display in visualisations and need to be considered when summarising the quantity and quality of evidence within- and across-indications. In this section we consider how to include these features in evidence visualisations to allow a judgement of the similarity of RTEs across indications and inform the decision of whether information across some or all indications can be combined to expand the evidence base.

## 3.1 Outcome data

In oncology trials, time-to-event data can be reported for different events of interest which can include time to reaching complete response, disease progression, or death. Other commonly reported outcomes include objective response rate (ORR) as well as OS and PFS. For regulatory and reimbursement authorities, OS at the end of trial follow-up is typically considered the outcome of primary interest but evidence on other outcomes such as PFS and ORR is often presented to accelerate drug approval and reimbursement decisions. Here, we will focus on OS and PFS, which are the time-to-event outcomes typically of primary interest for oncology HTA and commonly presented in published trial reports as hazard ratios (HRs) with uncertainty presented as 95% confidence intervals (CIs).

## 3.2 Data structure

Oncology trials typically report multiple outcomes over multiple time-points (interim or final), where evidence from earlier time-points is often used during drug appraisal by HTA bodies. Other trial characteristics that are important to display include differences between patient populations, the treatments administered (both intervention and comparators), or trial conduct across and within indications. Consideration of homogeneity and consistency within and across indications are important for making decisions about whether information can be combined both within and across indications

## 3.3 Quantity of evidence

It is also important to consider the quantity of the evidence available, and how it accumulates over time. In a multi-indication context, the quantity of evidence can be viewed in different ways, including the number of relevant studies within and across indications, the number of patients who took part in each trial, the number of events, and/or other relevant trial features such as the trial start and end dates (which inform trial duration). These features are also related to concepts of data maturity and uncertainty in RTE estimates, which are important considerations when making judgements about whether or not information should be borrowed across indications.

### 3.3.1 Maturity of evidence

Maturity of evidence on time-to-event data relates to how complete the trial is in terms of observing the event of interest in all individuals in the sample, at the time of reporting. This means that follow-up and censoring are both important in defining maturity. However, there is no single accepted metric of maturity. Often a quantification of length of follow-up (typically the median) is reported but in many trials it is unclear what the median refers to, and how it is interpreted.[21]

Evidence can be considered more mature if it is reported at a later timepoint.[22] However, this concept is not very useful when comparing the maturity of evidence across indications where a longer follow-up in one indication may not necessarily translate to more observed survival events if prognosis is more favourable than in other indications. A definition of maturity as the proportion of patients who experience an event relative to the total number of patients in the trial may provide more meaningful comparisons across indications.[23] For a given follow-up duration, Monnickendam et al.[24] calculated an index of completeness using information from digitised Kaplan-Meier (KM) curves, defined as the actual number of individuals that remain in follow-up as a proportion of the total number that could be expected to remain in the follow-up if data were entirely complete during a particular time interval. Although this provides a useful measure of data completeness which is comparable across indications, the data collection burden of digitizing all KM curves is considerable.

We define maturity as the number of events (OS or PFS) in each treatment arm at an interim or final timepoint divided by the total number of patients at the start of the trial.[23]

### 3.3.2 Uncertainty

When discussing the HRs for OS and PFS, it is important to also consider the uncertainty associated with the estimates from each trial. Firstly, we consider the width of the 95% CI (or credible interval, CrI), calculated as the difference between the upper and lower limits, where a smaller width indicates more precision in the estimate as a measure of uncertainty in the RTE. We considered a second measure of uncertainty, analogous to the coefficient of variation,[25] where the uncertainty was expressed relative to the magnitude of RTE and calculated on the log scale as $\frac{SE}{|\ln(HR)|}$. Here SE is the standard error of the ln(HR). The smaller the value of this ratio, the more precise (less uncertain) the estimate. The standard error can be useful to compare the precision of estimates across indications.

## 4 Methods

### 4.1 Evidence Synthesis

In HTA, meta-analyses are often conducted to pool results of multiple studies within the single, target, indication, to estimate the overall RTE.[26, 27] Common (also known as fixed) or random effects models can be used when RTEs estimated by the different studies are expected to be equal or heterogeneous, respectively. As some heterogeneity across studies within indication is expected, we will consider Bayesian random-effects meta-analysis models for pooling evidence within indications.[28, 29] We will consider Bayesian hierarchical meta-analysis models that allow borrowing of information on RTEs across indications.

A number of models have been proposed which differ in the level of sharing they allow across indications.[30] Here we extend standard meta-analysis models to the simplest borrowing models:

1) **Independent parameter (IP) model**, where the treatment effect for an indication is formed by the within-indication evidence only (no borrowing).
2) **Common parameter (CP) model**, which assumes that the treatment-effect is equal across indications so that a single/common effect is estimated for all indications (complete borrowing).
3) **Hierarchical meta-analysis (HMA) model**, where borrowing of information across indications is moderated by the between-indication heterogeneity. In a multi-indication context, this model is also referred to as panoramic meta-analysis [15-17].

For detailed specification of these models see Supplementary Section B.

We will implement these models using a cumulative meta-analysis[31, 32] framework to explore the change in estimated treatment effects over time. As the available evidence-base evolves a new meta-analysis is conducted every time a study reports its final outcome so that results include all evidence available at that point in time. Depending on the meta-analysis model used, cumulative meta-analyses could be only within-indication (IP model) or include evidence from other indications (CP and HMA models). We will consider different ways of visualising the results of these 3 meta-analysis models.

## 4.2 Displaying evidence and synthesis results

When considering displays to visualise multi-indication oncology evidence, we looked for displays that would clearly show the key features of interest, would be easy to understand and provide a visual indication of whether evidence is exchangeable across indications. In this section we propose a set of displays that can provide an overview of the evidence and how it evolves over time.

### 4.2.1 Visualising evidence accumulation over time

Timelines can be used to show how trial evidence accumulates over time, and the impact of accumulating evidence on estimated treatment effects.

A simple timeline (or time trend[33]) plot, with time represented on the horizontal axis, can be used to display the beginning and end of trials, as well as any interim timepoints when outcomes are reported. Timeline plots for each indication presented on the same display allow visualisation of the accumulation of evidence across all indications over time. These plots can be extended to emphasise other features such as the quantity (e.g. represented by the size of the trials), maturity, and uncertainty of the evidence at each time point.

### 4.2.2 Visualising outcome data

Traditionally, results of trials and the pooled estimates generated from a meta-analysis have been visually presented as forest plots[33, 34] that display point estimates as circles or squares, and their corresponding 95% CIs or CrIs as a line between the lower and upper bounds. Forest plots have been criticised for giving the false perception that all points within the interval are supported equally by the evidence.[35] Outcome data can instead be presented using density plots which provide an overview of the complete distribution instead of focusing on the point estimate and the corresponding 95% (or other) intervals. There is also less focus on the implications of statistical significance suggested by CIs as data are presented as a 'continuum of probability'.[36]

Ridgeline plots[37] can be used to display differences in densities between different groups, where distributions are represented as partially overlapping density plots that share a common scale on the horizontal axis. They are particularly useful when representing a large number of groups where separate plots might take up too much space and there is a clear pattern (e.g. rankings or ordering) to represent across time. In a multi-indication context, ridgeline plots can be constructed for each indication, displaying the density of the final reported relative effect measure (assuming a normal distribution, for example) for all reported outcomes.

### 4.2.3 Visualising synthesis results

Ridgeline plots can also be extended to display the results of cumulative meta-analyses performed for a single synthesis model, as well as to compare how pooled treatment effects differ for different evidence sharing models.

Violin plots[38] were proposed as a modification to the box-plot to show the underlying density together with the summary statistics. The density is mirrored across a central line where summary statistics can be depicted. Split violin plots are a variation of violin plots where two different densities can be plotted on each side of the central line, making it easier to compare distributions across two outcomes or from different analyses.[39] In a multi-indication context split-violin plots can be used to compare the OS and PFS estimated by different synthesis models across indications.

## 5 Results: evidence mapping in the bevacizumab case study

We created the plots discussed in this section using R[40] version 4.4.1. The meta-analyses conducted in Section 5.3 were conducted in R[40] using the R2OpenBUGS[41] package adapting the code developed in Singh et al.[30] We created the density plots for the results of the cumulative meta-analysis by directly plotting the output from the Markov Chain Monte Carlo (MCMC) simulations. However, the dataset

## 5.1 Displaying the evolution of evidence

The time summary plot summarising the bevacizumab evidence base is presented in Figure 2. Indications are presented in chronological order, starting with the indication with the earliest trial start date, colorectal cancer, at the top. The start of each trial is depicted by a small vertical line. We have denoted time points where interim and final HRs for OS and PFS were reported by a circle and a cross, respectively. Not all studies reported a HR for OS.

A horizontal line, depicting the duration of the trial, is used to join all outcome reporting timepoints. Differences in comparator treatments used in each RCT can be highlighted by using different line-types and colours. Most studies included in our dataset compared bevacizumab in combination with chemotherapy to chemotherapy alone. Following clinical advice, we decided not to differentiate between the different chemotherapy regimens as the treatment effect of bevacizumab would be unlikely to differ across different chemotherapies. Therefore, all studies where bevacizumab was compared in addition to chemotherapy are presented as black lines, and all other studies are shown as grey lines with details of the comparator added to the plot (Figure 2).

In Figure 3 we present examples of modified timeline plots, displaying other important data features using the NSCLC panel as an example. Complete versions of each modified timeline plot, showing all indications are included in Supplementary Section C (Figures S1-S5).

In Figure 3(b), the start point of each trial is depicted as a square, weighted according to the trial overall sample size, where the size of the square increases with an increase in the number of patients in the trial.

In Figure 3(c), we display the uncertainty in both PFS and OS, defined as the width of the 95% CI. The uncertainty in PFS and OS are shown as differently coloured circles: OS is depicted in black and PFS is represented in orange. Larger circles indicate greater precision. Modified timeline plots where uncertainty is defined as $\frac{SE}{|\ln(HR)|}$ are also included in the Supplementary Section C (Figure S5), where larger circles again indicate increased precision.

In Figure 3(d), we present the maturity of evidence for OS at each reporting time point. Circles for both treatment arms are weighted according to the magnitude of the maturity (described in Section 3.3.1). Black circles are used to represent the maturity of bevacizumab and red circles the maturity of the comparators. Larger circles represent more mature data i.e. where the proportion of patients who experience an event relative to the total number of patients in the trial is largest. Crosses are left to indicate points at which OS was reported but the measure of maturity could not be calculated.

In Figure 2, we can see that of all cancer types, trials have been conducted over the longest period of time (18 years) in NSCLC. The earliest trial (E4599) started in 2001 and the last trial, NEJ026, ended in 2019. Even in indications where more trials were conducted (e.g. colorectal and breast cancer), the trial period was shorter. In Figure 3, we can see that the first two trials (E4599 and AVAiL) were the largest trials and were also the only trials that allowed us to observe the maturity of the OS evidence. Due to the sparsity of data available to calculate the measures of maturity, it is difficult to comment on maturity within this indication. From Figure 3(c), we can see that both the OS and PFS reported for IMpower150 were not very precise and that for JO25567 and NEJ026 the reported OS was less precise than the PFS. Looking at the timeline plots, we can also see how treatments change over time - the later trials compare the effectiveness of bevacizumab to targeted therapies instead of just chemotherapy.

When choosing how to weight the circles, it is important to consider that there is a limit to how circles of different diameters appear distinct to the naked eye, and that if there are any extreme values of uncertainty/maturity it may become harder to differentiate between the less extreme values.

As our plots looked at a long period of time, when trial dates were too close together markers on the timeline plots tended to overlap, making it hard to distinguish between them. When this happened, we added an arbitrary gap of 2 months between two reporting points to improve visibility of these points in the plots. This gap in time was not incorporated into any other displays or in the syntheses.

The timeline plots show that licensing and technology appraisals occur shortly after indication-specific trials have reported results. The trials conducted did not always report OS and PFS at the same timepoints, with PFS results typically being reported earlier and therefore used more often to support HTA.

The timeline plots displaying evidence maturity are limited by the fact that very few trials report the number of events observed at a given timepoint, leaving us unable to examine the maturity of evidence across trials in a meaningful way. For example, in the NSCLC panel shown in Figure 3(d), while OS was reported at six time points, the number of events was reported only twice (where the maturity circles are shown). In the timeline plot displaying uncertainty, at the same timepoint, PFS estimates were generally more precise compared to the OS estimates.

The timeline plots shown in this section can be further extended by including markers to depict key events such as when the drug became available, drug licensing, and when a drug becomes the standard of care. In Singh et al.[30] timeline plots were extended to show every time bevacizumab was appraised by NICE.

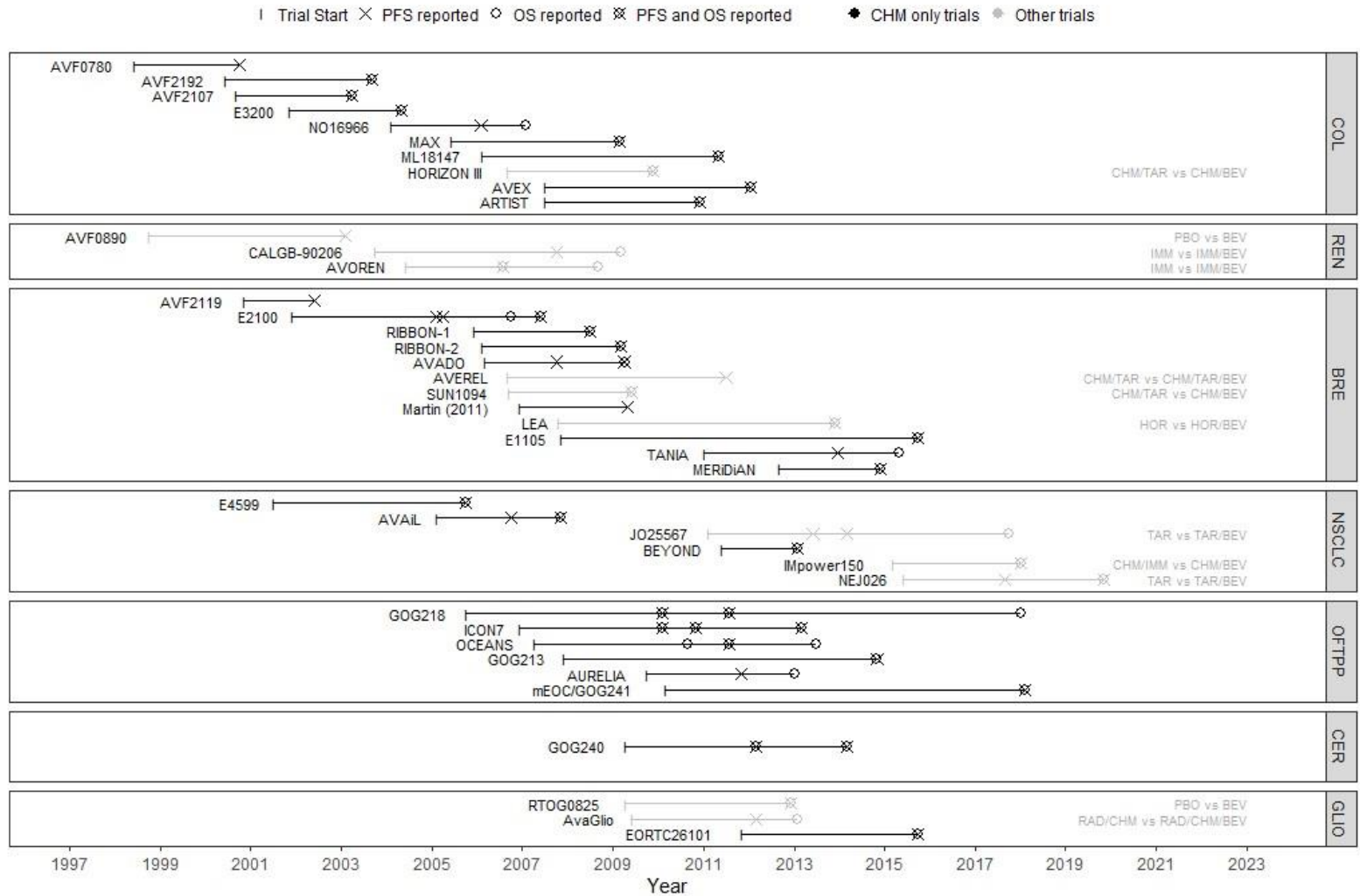

*Figure 2. Simple timeline plot of all licensed indications for bevacizumab*

**Abbreviations:** BEV, bevacizumab; BRE, breast cancer; CER, cervical cancer; CHM, chemotherapy; COL, colorectal cancer; GLIO, glioblastoma; HOR, hormonal therapy; IMM, immunotherapy; NSCLC, non-small cell lung cancer; OFTPP, ovarian, fallopian tube and primary peritoneal cancer; OS, overall survival; PBO, placebo; PFS, progression-free survival; RAD, radiotherapy; REN, renal cell carcinoma; TAR, targeted therapy.

*Figure 3. Timeline plots for NSCLC where the simple timeline plot (a) is presented with modified timeline plots showing: (b) start points weighted according to sample size of trial (c) the uncertainty in OS and PFS, measured as the width of the 95% CI, and (d) the maturity of the OS evidence*

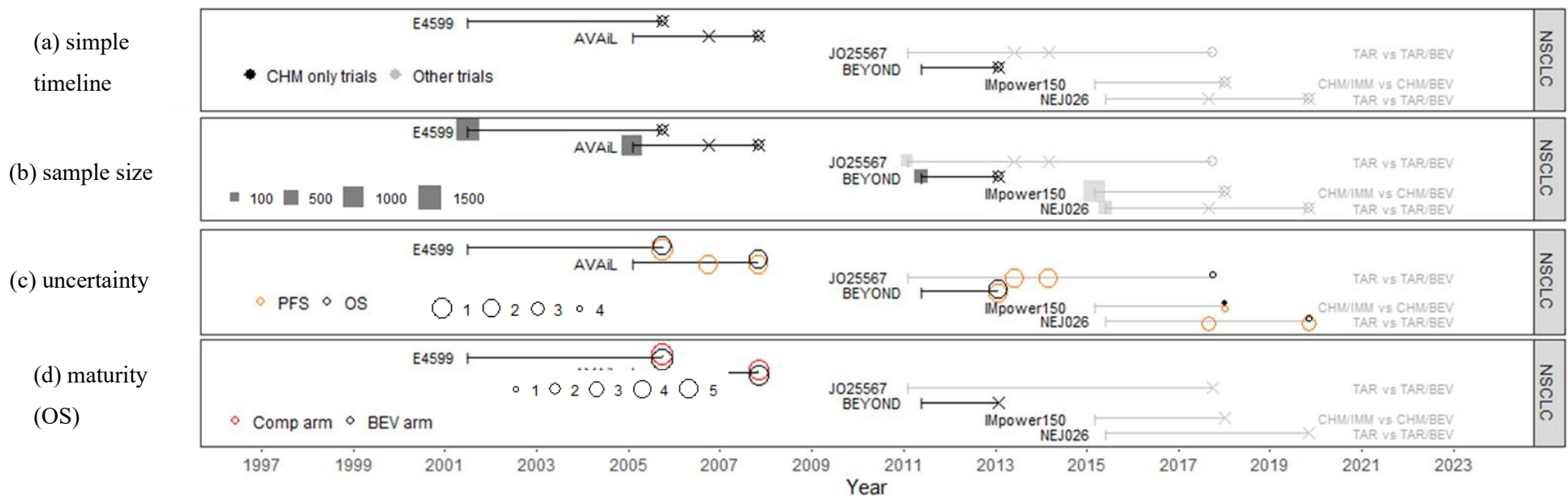

**Key for circle size:**

**(c) Uncertainty**: The circles in the legend have the following uncertainty values (calculated as the width of the CI) **1**:less than 0.25, **2**: 0.26 to 0.45, **3**: 0.46 to 0.65, **4**: 0.66 and over. For extreme values of uncertainty (defined as an uncertainty of more than 1.00), the uncertainty is represented by a point in the relevant colour.

**(d) Maturity**: The circles in the legend have the following maturity values (calculated as the proportion of events/total patients) **1**:less than 0.25, **2**: 0.26 to 0.40, **3**: 0.41 to 0.55, **4**: 0.56 to 0.70, **5:** 0.71 and over.

**Abbreviations:** BEV, bevacizumab; CHM, chemotherapy; CI, confidence interval; Comp, comparator; HOR, hormonal therapy; IMM, immunotherapy; NSCLC, non-small cell lung cancer; OS, overall survival; PBO, placebo; PFS, progression-free survival; TAR, targeted therapy.

## 5.2  Displaying relative effects

Ridgeline plots that show the accumulation of trial evidence for each indication over time are presented in Figure 4. These plots show the density of the final reported lnHRs for OS and PFS, assuming a normal distribution with variance calculated from the reported 95% CIs. The vertical axis shows the year outcomes were reported. Trials may not report all outcomes at the same time; for example, in colorectal cancer, trial NO16966 reported OS and PFS a year apart.

The ridgeline plots in Figure 4 show that for each outcome (PFS and OS), the curves overlap within and across indications, suggesting that the treatment effect of bevacizumab is similar across indications, although there is some heterogeneity between studies within indications.

The ridgeline plots for some indications (i.e. colorectal, breast and ovarian cancers) in Figure 4 are difficult to understand as many trials were conducted around the same time. For cluttered ridgeline plots, an alternative is to organise plots by effect size. An example of these plots can be seen in Supplementary Section C, Figure S6, where the ridgeline plots for all indications are ordered by decreasing OS. These plots allow us to compare the final reported OS and PFS within- and between-indications without considering when the trials were conducted.

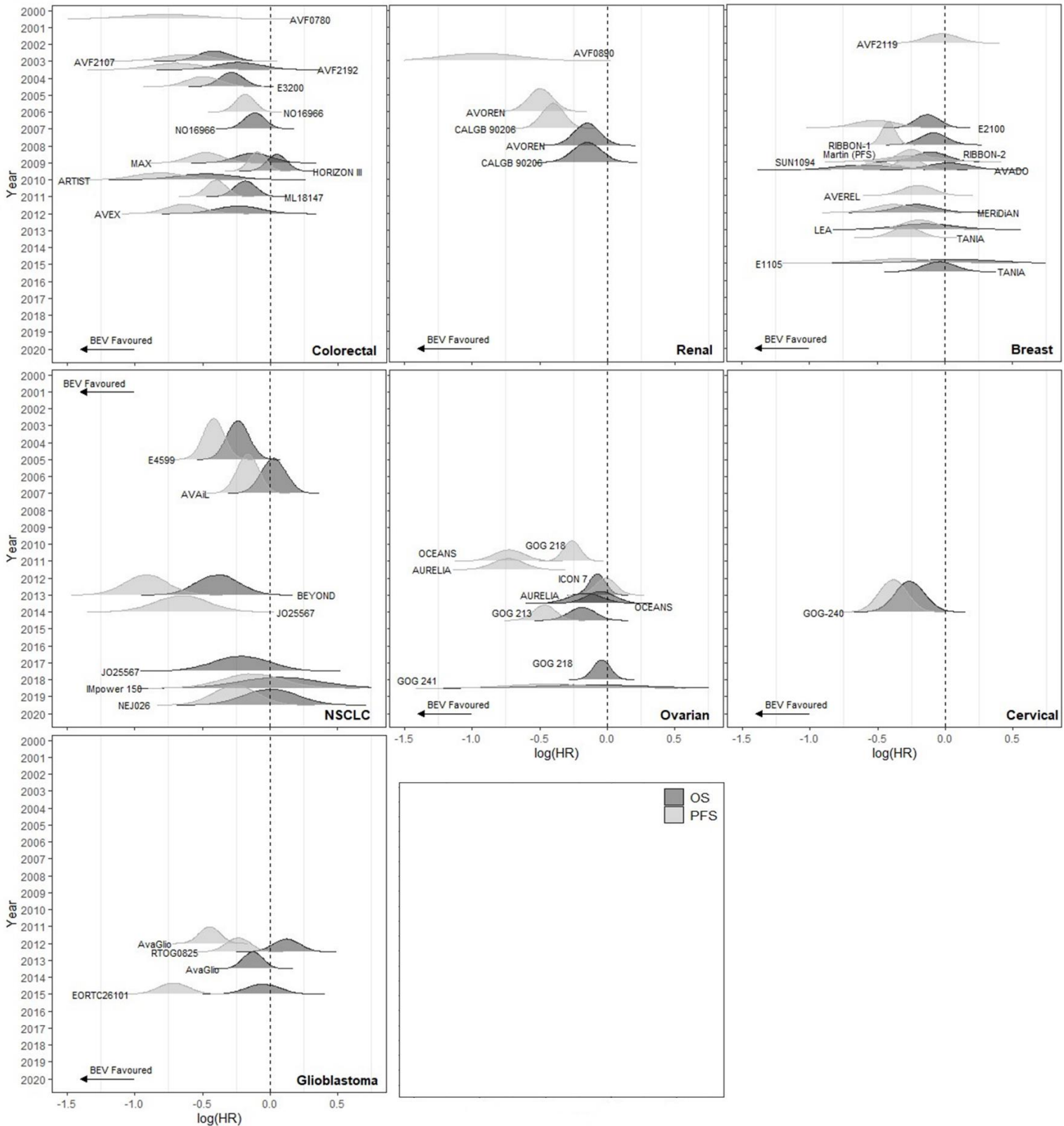

*Figure 4. Ridgeline plots for all licensed indications for bevacizumab. The legend for outcome type is included in the final panel.*

**Abbreviations:** BEV, bevacizumab; BRE, breast cancer; CER, cervical cancer; COL, colorectal cancer; GLIO, glioblastoma; HR, hazard ratio; NSCLC, non-small cell lung cancer; PFS, progression-free survival; OFTPP, ovarian, fallopian tube and primary peritoneal cancer; OS, overall survival; REN, renal cell carcinoma

.

## 5.3 Displays of synthesis results

In this section we describe the visualisations generated from the results of the cumulative meta-analyses. Detailed results for the meta-analyses, including estimated treatment effects, heterogeneity and model fit are provided in Supplementary Material Section B-II.

### 5.3.1 Ridgeline Plots

The evidence accumulation ridgeline plots in Figure 4 show that relative effects (on the log scale) for PFS and OS both are similar across indications, suggesting support for sharing evidence (borrowing information on the treatment effect) across indications.

The extended ridgeline plot in Figure 5 shows the results of cumulative meta-analysis using the IP model which synthesises evidence within each indication. The vertical axis represents time, and at each timepoint where a trial has reported a final outcome (in this case, OS), two density curves are plotted. The first (depicted in light gray) is the density for the lnHR reported in the study at that timepoint. The second (depicted as dark gray) is the density of the final relative effect measure for the cumulative meta-analysis conducted at that time-point, using all the evidence available up until that timepoint. The curves have been labelled with the name of the new trial being included.

The extended ridgeline plot can show the accumulation of evidence within an indication and how the pooled treatment effect changes with the inclusion of more evidence. The plots demonstrate that for all indications, as more evidence is added to the meta-analysis, the peak of the curve gets more pronounced, indicating an increase in the precision of the estimated RTE.

In this case-study, once 3 studies have been included in a cumulative analysis, the magnitude of the treatment effect (i.e. the position of the midpoint) stays largely consistent, and estimates become only slightly more precise (i.e. the spread of the distribution becomes slightly narrower). This may be due to the level of heterogeneity between the studies within-indication, which means additional studies will have little impact on the mean but can still have some impact on the precision.

Extended ridgeline plots can also be used to compare how pooled treatment effects for an indication differ using different cumulative meta-analysis models by super-imposing the three posterior distributions onto each other, as shown in Supplementary Section C Figures S7 and S8. These plots can demonstrate how treatment effects evolve over time, and how they differ according to different evidence sharing assumptions. For each indication, a new meta-analysis is conducted every time a study reports a new final outcome, but depending on the model used (IP, CP or HMA), only within-indication evidence or all available evidence from all indications at that timepoint are included in the meta-analysis. For both OS (Figure S7) and PFS (Figure S8), we can see that for all indications, the density for the CP model is shown as having the highest peak. This is what we would expect as this model includes the strongest sharing assumption across indications, increasing the precision of the estimates.

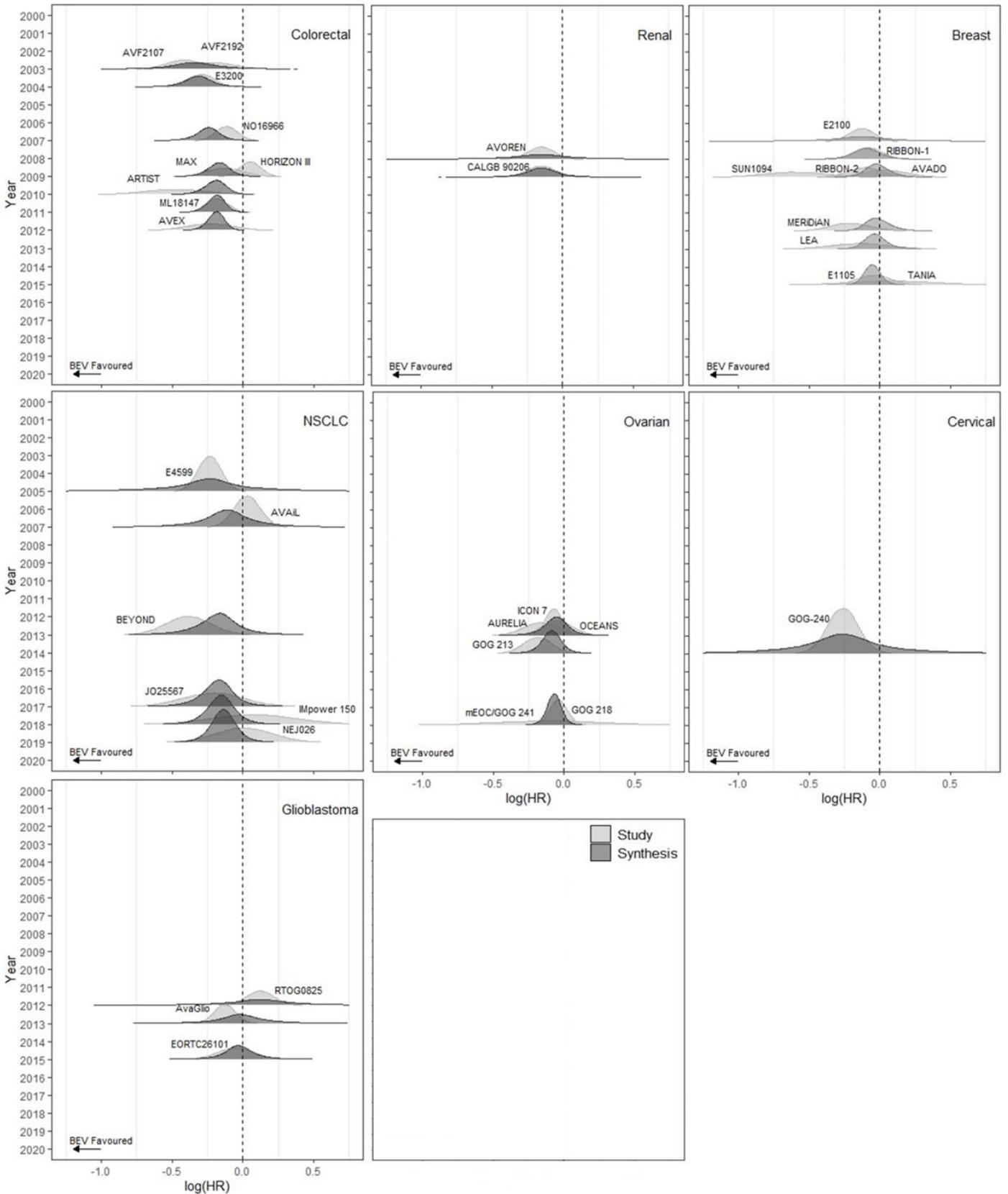

*Figure 5. Synthesis ridgeline plots for all licensed bevacizumab indications for overall survival.* The legend for the distribution is included in the final panel.

**Abbreviations:** BEV, bevacizumab; BRE, breast cancer; CER, cervical cancer; COL, colorectal cancer; GLIO, glioblastoma; HR, hazard ratio; NSCLC, non-small cell lung cancer; OFTPP, ovarian, fallopian tube and primary peritoneal cancer; REN, renal cell carcinoma.

### 5.3.2 Split-violin Plots

Split-violin plots can be used to display the impact of the three different models comparatively across indications (Figure 6). The lnHRs for OS and PFS, estimated after the publication of the final results for the last trial in each indication, are presented in these plots. The box-plots in the split violins highlight that for all indications the results of the synthesis models are largely consistent. These plots, like the ridgeline plots (Figures S7 and S8 in Supplementary Section C), show that the results from the IP model have the least precision which is consistent with the assumption made in the model where only indication-specific evidence is included in the synthesis.

*Figure 6. Split-violin plots comparing OS and PFS results for the three synthesis models*

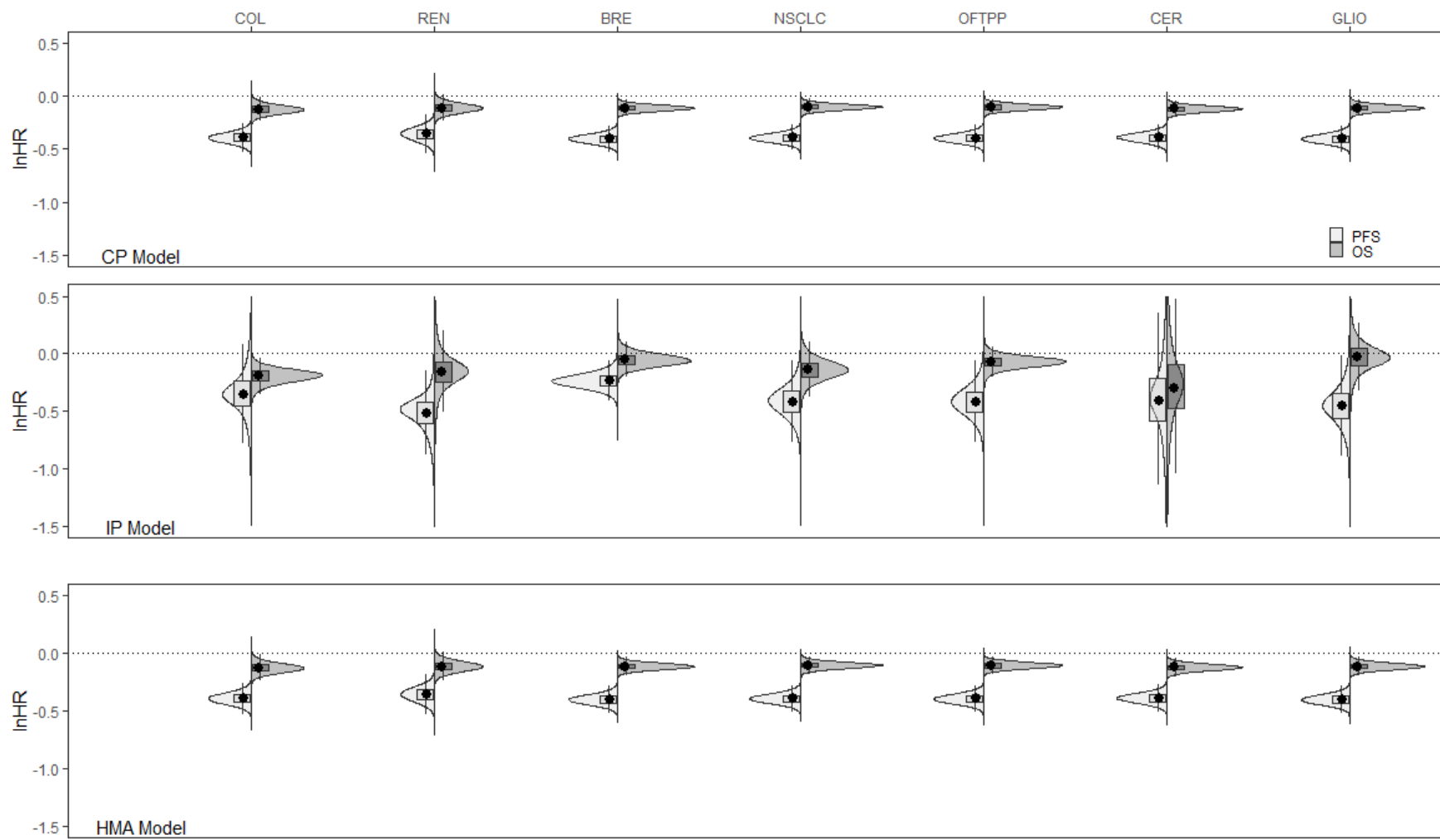

**Abbreviations:** BRE, breast cancer; CER, cervical cancer; COL, colorectal cancer; CP, common parameter; GLIO, glioblastoma; HR, hazard ratio; IP, independent parameter; NSCLC, non-small cell lung cancer; PFS, progression-free survival; HMA, hierarchical meta-analysis; OFTPP, ovarian, fallopian tube, and primary peritoneal cancer, OS, overall survival; REN, renal cell carcinoma

# 6 Discussion

With an increase in the availability of multi-indication therapies, there is a growing interest in approaches for the evaluation of these technologies. Understanding the complex evidence-base is imperative to developing these methods and evaluating assumptions. Effective visualisation techniques can be useful for better communicating and understanding the evidence-base. The visualisation methods discussed in this paper can be modified to capture features of the evidence that are of interest for analysts and policymakers. The plots presented here (timeline, ridgeline, split-violin) can be adapted in simple ways to explore other contexts where the sharing of information is of interest- including the use of direct and indirect evidence, or multiple drugs of the same class used in the same indication. The methods presented in this paper can also be extended to show the results for more complex mixture models[30]. An empirical assessment of whether the assumptions made for the evidence synthesis modelling were appropriate are beyond the scope of this work, but they are discussed in detail in Singh et al.[30]

We only considered licensed indications in our case study as our aim was to judge similarity across comparable indications. We expect relative effects in non-licensed indications to be different from licensed indications as a reason for no license may be related to a lack of efficacy. However, it may be that there are other reasons for no license- this should be discussed with topic experts on a case-by-case basis when considering which indications to include. Evidence displays could help structure the discussions and make these judgements.

The results of the synthesis models indicated that the simple CP model where there was maximum sharing of evidence provided the most precise results. However, the 'lumping' together of all trials across indications may add bias, as this strong assumption is unlikely to be valid across all indications. A discussion of the trade-off between precision and bias is needed. As in any synthesis, expert opinion on the plausibility of assumptions and formal statistical checks for model fit should be considered. The plots displaying synthesis results compare models with different assumptions; however, they do not provide any clarity on whether the assumptions made are correct. Plots that display the data can help inform judgements on which assumptions may be most appropriate.

In our illustrative case-study, while our aim was to identify as many RCTs comparing bevacizumab as possible, due to time and resource constraints the searches conducted were not comprehensive. Therefore, the evidence-base presented here may not be exhaustive.

The lack of evidence available and the inconsistent reporting of useful evidence measures prevented us from visualising some key features of the evidence effectively. In particular, since so few studies reported the number of events observed during a trial, we were unable to compare the maturity of evidence across all trials. This could be improved by using other measures of maturity or by digitizing Kaplan-Meier curves, where presented.

For the visualisation of RTEs, we used density instead of the point estimate and corresponding 95% CI, an approach that was well-received and understood by all clinical co-authors. However, the ridgeline plots presented here can be developed further to provide more than a general impression of the evidence available, especially when there is a lot of evidence within a short period of time. The ridgeline plots for some indications (i.e. colorectal, breast and ovarian cancers) in Figure 4 are difficult to interpret as many trials were conducted around the same time. A potential extension to these ridgeline plots is to make them dynamic so that stakeholders are able to query the data further by, for example, clicking on particular regions of interest.

The displays presented here could be extended to visualise other features not addressed in this work including subgroups, differences in study design, the quality of studies, and statistical considerations such as non-proportional hazards, cross-over adjustments and stratification. There is also a need to look at additional relevant outcomes that may be used in HTA, such as the response rate. However, incorporating new outcomes introduces additional challenges in the visualisation of evidence. The joint presentation of outcomes on different scales will require modifications to the plots and may not always be useful. In our examples, treatment effects on both outcomes were on the lnHR scale and presenting ln(Odds Ratios) for response on the same plot as the lnHRs for OS and PFS would require modifications to accurately represent the differences in scales between the different outcomes.

Bevacizumab was used as a case-study due to its many licensed indications. However, it may be less complex than typical oncology drugs in that the treatment effect is likely to be exchangeable when given together with different background therapies (i.e., it is less likely to be modified by interactions with background treatments than other oncology drugs). This is because bevacizumab specifically is deemed to administer its effect by its interaction on the stromal environment to the cancer as opposed to the tumour cells themselves and there is likely to be more consistency between tumours in relation to this. Therefore, while clinical heterogeneity did not appear to matter for our example, it may be important to consider for other multi-indication drugs. In addition to heterogeneity between-indications, heterogeneity may also exist within-indications making it necessary to look at trial designs, subgroups, lines of treatment and comparator treatments. This can be accomplished by tailoring the plots presented here to highlight causes for heterogeneity between and within-indications.

**Acknowledgements**

Financial support for this study was provided entirely by a grant from the Medical Research Council (grant No. MRC MR/W021102/1).

# Supplementary Material

# A: Study identification and data extraction

## A-I: Identification of studies

**Table S1.** Bevacizumab trials that were identified.

| Study | Publications | Control | Comparator[‡] |
|---|---|---|---|
| *Breast Cancer* | | | |
| AVF2119 | Miller (2005)[1] | Capecitabine | Capecitabine + Bevacizumab |
| E2100 | Miller (2007)[2]; Cameron (2008)[3] | Paclitaxel | Paclitaxel + Bevacizumab |
| RIBBON-1 | Robert (2011)[4] | Capecitabine | Capecitabine + Bevacizumab |
| | | Taxane/ Anthracycline | Taxane/Anthracycline + Bevacizumab |
| RIBBON-2 | Brufsky (2011)[5] | Chemotherapy | Chemotherapy + Bevacizumab |
| AVADO | Miles (2010)[6]; Miles (2013)[7] | Docetaxel | Docetaxel + Bevacizumab (15mg/kg) |
| AVEREL | Gianni (2013)[8] | Docetaxel +Trastuzumab | Docetaxel + Trastuzumab + Bevacizumab |
| SUN1094 | Robert (2011)[9] | Paclitaxel + Sunitinib | Paclitaxel + Bevacizumab |
| Martin (2011) | Martin (2011)[10] | Paclitaxel + Placebo | Paclitaxel+ Bevacizumab |
| LEA | Martin (2015)[11] | Endocrine therapy | Endocrine therapy + Bevacizumab |
| E1105 | Artega (2012)[12]; Clinicaltrials.gov[13] | Chemotherapy + Placebo | Chemotherapy + Bevacizumab |
| TANIA | Von Minckwitz (2014)[14]; Vrdoljak (2016)[15] | Chemotherapy | Chemotherapy + Bevacizumab |
| MERiDiAN | Miles (2017)[16] | Placebo + Paclitaxel | Bevacizumab + Paclitaxel |
| *Cervical Cancer* | | | |
| GOG 240 | Tewari (2014)[17]; Tewari (2017)[18] | Chemotherapy | Chemotherapy + Bevacizumab |
| *Colorectal Cancer* | | | |
| AVF0780 | Kabbinavar (2003)[19] | FL | FL + Bevacizumab (5 mg/kg) |
| AVF2192 | Kabbinavar (2005)[20] | FL | FL + Bevacizumab |
| AVF2107 | Hurwitz (2004)[21] | IFL + Placebo | IFL + Bevacizumab |
| E3200 | Giantonio (2007)[22] | FOLFOX4 | FOLFOX4 + Bevacizumab |

| Study | Publications | Control | Comparator[‡] |
|---|---|---|---|
| NO16966 | Saltz (2008)[23]; Cassidy (2011)[24] | Chemotherapy | Chemotherapy + Bevacizumab |
| MAX | Tebbutt (2010)[25] | Capecitabine | Capecitabine + Bevacizumab |
| ML18147 | Bennouna (2013)[26]; Kubicka (2013)[27] | Chemotherapy | Chemotherapy + Bevacizumab |
| HORIZON III | Schmoll (2012)[28] | mFOLFOX6 + Cediranib | mFOLFOX6 + Bevacizumab |
| AVEX | Cunningham (2013)[29] | Capecitabine | Capecitabine + Bevacizumab |
| ARTIST | Guan (2011)[30] | IFL | IFL + Bevacizumab |
| *Glioblastoma* | | | |
| RTOG0825 | Gilbert (2014)[31] | Placebo | Placebo + Bevacizumab |
| AVAglio | Sandmann (2015)[32] | Radiotherapy/ Temozolomide | Radiotherapy/Temozolomide + Bevacizumab |
| EORTC26101 | Wick (2017)[33] | Lomustine | Lomustine + Bevacizumab |
| *NSCLC* | | | |
| E4599 | Sandler (2006)[34] | Carboplatin + Paclitaxel | Carboplatin + Paclitaxel + Bevacizumab |
| AVAiL | Reck (2009)[35]; Reck (2010)[36] | Cisplatin + Gemcitabine + Placebo | Cisplatin + Gemcitabine + Bevacizumab (15 mg/kg) |
| JO25567 | Seto (2014)[37]; Yamamoto (2021)[38] | Erlotinib | Erlotinib + Bevacizumab |
| BEYOND | Zhou (2015)[39] | Carboplatin + Paclitaxel + Placebo | Carboplatin + Paclitaxel + Bevacizumab |
| IMpower150 | Reck (2019)[40]; Socinski (2021)[41] | Carboplatin + Paclitaxel + Atezolizumab | Carboplatin + Paclitaxel + Bevacizumab |
| NEJ026 | Saito (2019)[42]; Kawashima (2022)[43] | Erlotinib | Erlotinib + Bevacizumab |
| *Ovarian, fallopian tube, and primary peritoneal cancer* | | | |
| GOG218 | Burger (2011)[44]; Tewari (2019)[45] | Carboplatin + Paclitaxel + Placebo | Carboplatin + Paclitaxel + Bevacizumab |
| ICON7 | Perren (2011)[46]; Oza (2015)[47] | Carboplatin + Paclitaxel | Carboplatin + Paclitaxel + Bevacizumab |
| OCEANS | Aghajanian (2012)[48]; Aghajanian (2015)[49] | Gemcitabine + Carboplatin + Placebo | Gemcitabine + Carboplatin + Bevacizumab |
| GOG213 | Coleman (2017)[50] | Carboplatin + Paclitaxel | Carboplatin + Paclitaxel + Bevacizumab |
| AURELIA | Pujade-Lauraine (2014)[51]; Bamias (2017)[52] | Chemotherapy | Chemotherapy + Bevacizumab |
| mEOC/GOG241 | Gore (2019)[53] | Carboplatin + Paclitaxel | Carboplatin + Paclitaxel + Bevacizumab |
| | | Oxaliplatin + Capecitabine | Oxaliplatin + Capecitabine + Bevacizumab |
| *Renal cell carcinoma* | | | |

| Study | Publications | Control | Comparator[‡] |
|---|---|---|---|
| AVF0890 | Yang (2003)[54] | Placebo | Bevacizumab (10 mg/kg) |
| CALGB-90206 | Rini (2008)[55]; Rini (2010)[56] | Interferon | Interferon + Bevacizumab |
| AVOREN | Escuidier (2007)[57]; Escuidier (2010)[58] | Interferon + Placebo | Interferon + Bevacizumab |
| *Gastrointestinal cancer*[†] | | | |
| AVATAR | Shen (2015)[59] | Capecitabine + Cisplatin + Placebo | Capecitabine + Cisplatin + Bevacizumab |
| AVAGAST | Ohtsu (2011)[60] | Capecitabine + Cisplatin + Placebo | Capecitabine + Cisplatin + Bevacizumab |
| *Lymphoma*[†] | | | |
| MAIN | Seymour (2014)[61] | R-CHOP* | R-CHOP + Bevacizumab |
| *Urothelial cancer*[†] | | | |
| CALGB-90601 | Rosenberg (2021)[62] | Cisplatin + Gemcitabine + Placebo | Cisplatin + Gemcitabine + Bevacizumab |
| *Prostate cancer*[†] | | | |
| CALGB-90401 | Kelly (2012)[63] | Docetaxel + Prednisone | Docetaxel + Prednisone + Bevacizumab |
| *Uterine cancer*[†] | | | |
| GOG250 | Hensley (2015)[64] | Gemcitabine + Docetaxel + Placebo | Gemcitabine + Docetaxel + Bevacizumab |

* R-CHOP consists of Rituximab, Cyclophosphamide, Doxorubicin, Vincristine, and Prednisone. † These cancer indications were not included in the work conducted in this paper. ‡Where a dose is specified for bevacizumab, that was the trial arm that data were extracted from when a trial looked at multiple doses of bevacizumab.

**Treatment abbreviations:** FL- leucovorin and fluorouracil; IFL-irinotecan, leucovorin and fluorouracil; FOLFOX4- oxaliplatin, leucovorin, and 5- fluorouracil; mFOLFOX6- modified FOLFOX6; R-CHOP-rituximab, cyclophosphamide, doxorubicin, vincristine and prednisone.

## A-II: Data Extraction Details

For all included studies we extracted relevant trial characteristics as well as outcome data.

<u>Trial characteristics:</u>

We extracted the trial location and number of centres in the trials, details on treatment regimens (including doses, frequency, and duration of treatment). For trials where different doses of bevacizumab were compared to each other as well as a comparator treatment, we only extracted evidence from the treatment arm that used the dose licensed for that particular indication. We also extracted the length of follow-up in each trial.

<u>Patient characteristics:</u>

We extracted the patient demographics including age, sex, ECOG performance score, and prior treatment history, noted where subgroup analysis had been conducted on different patient characteristics.

<u>Outcome Data:</u>

We extracted outcome data for overall survival (OS), progression-free PFS, and response. For OS and PFS, we extracted the reported hazard ratio (HR) and 95% CI and, where reported, the number of participants who experienced an event (i.e. progression or death). For PFS we also extracted how progression was assessed and where trials reported more than one method, Independent Reviewer Committee/Facility (IRC/IRF) was preferred over investigator assessment.

For response, we recorded the overall response rates (ORRs), and the number of patients who experienced complete or partial response (CR or PR); however we did not explore response as outcome in our visualisations.

The extracted data that were used in the figures and analyses that were conducted are reported in Table S2 (for OS) and Table S3 (for PFS)

**Table S2**. Extracted data for OS

| Trial | Publication | Cut-off date[†] | Randomised Patients | | Number of Events | | Hazard Ratio |
| --- | --- | --- | --- | --- | --- | --- | --- |
| | | | Control | Comparator | Control | Comparator | (95% CI) |
| *Colorectal cancer* | | | | | | | |
| AVF2192 | Kabbinavar (2005) | 01/09/2003 | 105 | 104 | NR | NR | 0.79 (0.56, 1.10) |
| AVF2107 | Hurwitz (2004) | 01/04/2003 | 411 | 402 | NR | NR | 066 (0.52, 0.84) |
| E3200 | Giantonio (2007) | 01/05/2004 | 291 | 286 | NR | NR | 0.75 (0.63, 1.89) |
| NO16966 | Saltz (2008) | 01/02/2007 | 701 | 699 | NR | NR | 0.89 (0.76, 1.03) |
| MAX | Tebbutt (2010) | 27/02/2009 | 156 | 157 | NR | NR | 0.88 (0.68, 1.13) |
| ML18147 | Bennouna (2013) | 01/05/2011 | 411 | 409 | NR | NR | 0.83 (0.71, 0.97) |
| HORIZON-III | Schmoll (2012) | 15/11/2009 | 709 | 713 | 239 | 247 | 1.05 (0.91, 1.22) |
| AVEX | Cunningham (2013) | 19/01/2012 | 140 | 140 | NR | NR | 0.79 (0.57, 1.09) |
| ARTIST | Guan (2011) | 01/12/2010 | 64 | 139 | NR | NR | 0.62 (0.41, 0.95) |
| *Renal cell carcinoma* | | | | | | | |
| CALGB-90206 | Rini (2010) | 01/03/2009 | 363 | 369 | NR | NR | 0.86 (0.73, 1.10) |
| AVOREN | Escudier (2007) | 01/08/2006 | 322 | 327 | 137 | 114 | 0.75 (0.58, 0.97) |
| | Escudier (2010) | 01/09/2008 | 322 | 327 | 224 | 220 | 0.86 (0.72, 1.04) |
| *Breast cancer* | | | | | | | |
| E2100 | Cameron (2008) | 01/10/2006 | 354 | 368 | NR | NR | 0.87 (0.72, 1.05) |
| | Miller (2007) | 01/06/2007 | 354 | 368 | NR | NR | 0.88 (0.74, 1.05) |
| RIBBON-1[‡] | Robert (2011) | 01/07/2008 | 206 | 409 | NR | NR | 0.85 (0.63, 1.14) |
| | | | 207 | 415 | NR | NR | 1.03 (0.77, 1.38) |
| RIBBON-2 | Brufsky (2011) | 01/03/2009 | 255 | 459 | 109 | 206 | 0.90 (0.71, 1.33) |
| AVADO | Miles (2010) & (2013) | 01/04/2009 | 241 | 247 | 133 | 131 | 1.03 (0.70, 1.33) |
| SUN1094 | Robert (2011) | 01/06/2009 | 242 | 243 | 52 | 32 | 0.55 (0.35, 0.86) |
| LEA | Martin (2015) | 01/12/2013 | 184 | 190 | 46 | 47 | 0.87 (0.58, 1.32) |
| E1105 | Clinical Trials Results | 01/10/2015 | 48 | 48 | NR | NR | 1.09 (0.61, 1.97) |

| Trial | Publication | Cut-off date† | Randomised Patients | | Number of Events | | Hazard Ratio (95% CI) |
|---|---|---|---|---|---|---|---|
| | | | Control | Comparator | Control | Comparator | |
| TANIA | Vrdoljak (2016) | 30/04/2015 | 247 | 247 | 156 | 163 | 0.96 (0.76, 1.21) |
| MERiDiAN | Miles (2017) | 30/11/2014 | 233 | 238 | 105 | 91 | 0.81 (0.61, 1.08) |
| *NSCLC* | | | | | | | |
| E4599 | Sandler (2005) | 01/10/2005 | 444 | 434 | 344 | 305 | 0.79 (0.67, 0.92) |
| AVAiL | Reck (2010) | 01/11/2007 | 347 | 351 | 240 | 242 | 1.03 (0.86, 1.23) |
| J025567 | Yamamoto (2021) | 01/10/2017 | 77 | 75 | NR | NR | 0.81 (0.53, 1.23) |
| BEYOND | Zhou (2015) | 27/01/2013 | 138 | 138 | NR | NR | 0.68 (0.50, 0.93) |
| IMpower150 | Reck (2019) | 01/01/2018 | 402 | 400 | NR | NR | 1.08 (0.60, 1.96) |
| NEJ026 | Kawashima (2022) | 01/11/2019 | 114 | 114 | NR | NR | 1.01 (0.68, 1.49) |
| *Ovarian, fallopian tube, and primary peritoneal cancer* | | | | | | | |
| GOG218 | Burger (2011) | 01/02/2010 | 625 | 623 | 156 | 138 | 0.92 (0.73, 1.15) |
| | Burger (2011) | 01/08/2011 | 625 | 623 | 298 | 269 | 0.89 (0.75, 1.04) |
| | Tewari (2019) | 01/01/2018 | 625 | 623 | NR | NR | 0.96 (0.85, 1.09) |
| ICON7 | Perren (2011) | 01/02/2010 | 764 | 764 | 130 | 111 | 0.81 (0.63, 1.04) |
| | Perren (2011) | 01/11/2010 | 764 | 764 | 200 | 178 | 0.85 (0.69, 1.04) |
| | Oza (2015) | 01/03/2013 | 764 | 764 | 352 | 362 | 0.99 (0.85, 1.14) |
| OCEANS | Aghajanian (2012) | 01/09/2010 | 242 | 242 | NR | NR | 0.75 (0.54, 1.05) |
| | Aghajanian (2012) | 01/08/2011 | 242 | 242 | NR | NR | 1.03 (0.79, 1.33) |
| | Aghajanian (2015) | 01/07/2013 | 242 | 242 | NR | NR | 0.95 (0.77, 1.18) |
| GOG213 | Coleman (2017) | 01/11/2014 | 337 | 337 | 214 | 201 | 0.83 (0.68, 1.01) |
| AURELIA | Pujade-Lauraine (2014) | 01/01/2013 | 182 | 179 | 136 | 128 | 0.85 (0.66, 1.08) |
| mEOC/GOG241‡ | Gore (2019) | 01/02/2018 | 13 | 11 | NR | NR | 1.47 (0.56, 3.84) |
| | | | 13 | 13 | NR | NR | 0.77 (0.29, 2.03) |
| *Cervical cancer* | | | | | | | |
| GOG240 | Tewari (2014) | 01/03/2012 | 225 | 227 | 140 | 131 | 0.71 (0.54, 0.95) |
| | Tewari (2017) | 01/03/2014 | 225 | 227 | 175 | 173 | 0.77 (0.62, 0.95) |

| Trial | Publication | Cut-off date† | Randomised Patients | | Number of Events | | Hazard Ratio (95% CI) |
|---|---|---|---|---|---|---|---|
| | | | Control | Comparator | Control | Comparator | |
| ***Glioblastoma*** | | | | | | | |
| RTOG0825 | Glibert (2014) | 01/10/2015 | 317 | 320 | 198 | 215 | 1.13 (0.93, 1.37) |
| AvaGlio | Chinot (2014) | 01/02/2013 | 463 | 458 | NR | NR | 0.88 (0.76, 1.02) |
| EORTC26101 | Wick (2017) | 01/10/2015 | 149 | 288 | 113 | 216 | 0.95 (0.74, 1.21) |

† Where studies only reported month and year for the data cut-off, we assumed that this was the first of the month. ‡Where studies reported multiple two-arm (chemotherapy vs. chemotherapy +bevacizumab) comparisons, both were included as long as there was no overlap in patients.

**Abbreviations:** CI, confidence interval; NR, not reported; NSCLC, non-small cell lung cancer

**Table S3.** Extracted data for PFS

| Trial | Publication | Cut-off date† | Assessment Method | Randomised Patients | | Number of Events | | Hazard Ratio (95% CI) |
|---|---|---|---|---|---|---|---|---|
| | | | | Control | Comparator | Control | Comparator | |
| *Colorectal cancer* | | | | | | | | |
| AVF0780 | Kabbinavar (2003) | 01/10/2000 | IRF | 36 | 35 | 26 | 22 | 0.46 (0.27, 0.79) |
| AVF2192 | Kabbinavar (2005) | 01/09/2003 | IRF | 105 | 104 | NR | NR | 0.50 (0.34, 0.73) |
| AVF2107 | Hurwitz (2004) | 01/04/2003 | IRC | 411 | 402 | NR | NR | 0.54 (0.37, 0.78) |
| E3200 | Giantonio (2007) | 01/05/2004 | INV | 291 | 286 | NR | NR | 0.61 (0.48, 078) |
| NO16966 | Saltz (2008) | 01/02/2006 | INV | 701 | 699 | NR | NR | 0.83 (0.72, 0.95) |
| MAX | Tebbutt (2010) | 27/02/2009 | NR | 156 | 157 | NR | NR | 0.62 (0.49, 0.79) |
| ML18147 | Bennouna (2013) | 01/05/2011 | INV | 411 | 409 | NR | NR | 0.67 (0.58, 0.78) |
| HORIZON-III | Schmoll (2012) | 15/11/2009 | NR | 709 | 713 | 471 | 453 | 0.91 (0.80, 1.03) |
| AVEX | Cunningham (2013) | 19/01/2012 | NR | 140 | 140 | NR | NR | 0.53 (0.41, 0.69) |
| ARTIST | Guan (2011) | 01/12/2010 | INV | 64 | 139 | NR | NR | 0.44 (0.31, 0.63) |
| *Renal cell carcinoma* | | | | | | | | |
| AVF0890 | Yang (2003) | 01/02/2003 | NR | 40 | 39 | NR | NR | 0.39 (0.23, 0.68) |
| CALGB-90206 | Rini (2008) | 01/10/2007 | INV | 363 | 369 | NR | NR | 0.67 (0.57, 0.79) |
| AVOREN | Escudier (2007) | 01/08/2006 | INV | 322 | 327 | 275 | 230 | 0.61 (0.51, 0.73) |
| *Breast Cancer* | | | | | | | | |
| AVF2119 | Miller (2005) | 01/06/2002 | IRC | 230 | 232 | NR | NR | 0.98 (0.77, 1.25) |
| E2100 | Cameron (2008) | 01/02/2005 | INV | 354 | 368 | 244 | 201 | 0.42 (0.34, 0.52) |
| | Cameron (2008) | 01/04/2005 | IRC | 354 | 368 | 184 | 173 | 0.48 (0.33, 0.69) |
| | Miller (2007) | 01/06/2007 | NR | 326 | 347 | 308 | 316 | 0.60 (0.44, 0.81) |
| RIBBON-1‡ | Robert (2011) | 01/07/2008 | IRC | 206 | 409 | NR | NR | 0.69 (0.56, 0.84) |
| | | | | 207 | 415 | NR | NR | 0.64 (0.52, 0.80) |
| RIBBON-2 | Brufsky (2011) | 01/03/2009 | INV | 255 | 459 | 184 | 372 | 0.78 (0.64, 0.93) |
| AVADO | Miles (2013) | 01/10/2007 | INV | 241 | 247 | NR | NR | 0.61 (0.48, 0.78) |

| Trial | Publication | Cut-off date† | Assessment Method | Randomised Patients | | Number of Events | | Hazard Ratio |
|---|---|---|---|---|---|---|---|---|
| | | | | Control | Comparator | Control | Comparator | (95% CI) |
| | Miles (2010) | 01/04/2009 | INV | 241 | 247 | 219 | 220 | 0.77 (0.64, 0.93) |
| AVEREL | Gianni (2013) | 30/06/2011 | INV | 208 | 216 | 154 | 153 | 0.82 (0.65, 1.02) |
| SUN1094 | Robert (2011) | 01/06/2009 | NR | 242 | 243 | 89 | 70 | 0.61 (0.44, 0.85) |
| Martin (2011) | Martin (2011) | 01/05/2009 | IRC | 94 | 97 | 15 | 9 | 0.79 (0.53, 1.17) |
| LEA | Martin (2015) | 01/12/2013 | NR | 184 | 190 | 135 | 128 | 0.83 (0.65, 1.06) |
| E1105 | Clinical Trials | 01/10/2015 | NR | 48 | 48 | NR | NR | 0.73 (0.43, 1.23) |
| TANIA | von Minckwitz (2014) | 20/12/2013 | INV | 247 | 247 | 203 | 204 | 0.75 (0.61, 0.93) |
| MERiDiAN | Miles (2017) | 30/11/2014 | INV | 233 | 238 | 168 | 152 | 0.68 (0.51, 0.91) |
| *Non-small cell lung cancer* | | | | | | | | |
| E4599 | Sandler (2005) | 01/10/2005 | NR | 444 | 434 | 405 | 374 | 0.66 (0.57, 0.77) |
| AVAiL | Reck (2019) | 01/10/2006 | INV | 347 | 351 | NR | NR | 0.82 (0.68, 0.98) |
| | Reck (2010) | 01/11/2007 | INV | 347 | 351 | NR | NR | 0.85 (0.73, 1.00) |
| J025567 | Seto (2014) | 01/06/2013 | IRC | 77 | 75 | 57 | 46 | 0.54 (0.36, 0.79) |
| | Yamamoto (2021) | 01/03/2014 | INV | 77 | 75 | NR | NR | 0.52 (0.35, 0.76) |
| BEYOND | Zhou (2015) | 27/01/2013 | INV | 138 | 138 | NR | NR | 0.40 (0.29, 0.54) |
| IMpower150 | Reck (2019) | 01/01/2018 | INV | 402 | 400 | NR | NR | 0.88 (0.56, 1.37) |
| NEJ026 | Saito (2019) | 01/09/2017 | IRC | 114 | 114 | NR | NR | 0.61 (0.42, 0.88) |
| | Kawashima (2022) | 01/11/2019 | INV | 114 | 114 | NR | NR | 0.77 (0.56, 1.07) |
| *Ovarian, fallopian tube, and primary peritoneal cancer* | | | | | | | | |
| GOG218 | Burger (2011) | 01/02/2010 | NR | 625 | 623 | NR | NR | 0.72 (0.63, 0.82) |
| | Burger (2011) | 01/08/2011 | NR | 625 | 623 | NR | NR | 0.77 (0.68, 0.87) |
| ICON7 | Perren (2011) | 01/02/2010 | INV | 764 | 764 | 392 | 367 | 0.81 (0.70, 0.94) |
| | Perren (2011) | 01/11/2010 | INV | 764 | 764 | 392 | 367 | 0.87 (0.77, 0.99) |
| | Oza (2015) | 01/03/2013 | INV | 764 | 764 | 526 | 554 | 0.93 (0.83, 1.05) |
| OCEANS | Aghajanian (2012) | 01/08/2011 | INV | 242 | 242 | 187 | 151 | 0.48 (0.39, 0.61) |
| GOG213 | Coleman (2017) | 01/11/2014 | INV | 337 | 337 | NR | NR | 0.63 (0.53, 0.74) |

| Trial | Publication | Cut-off date† | Assessment Method | Randomised Patients | | Number of Events | | Hazard Ratio (95% CI) |
|---|---|---|---|---|---|---|---|---|
| | | | | Control | Comparator | Control | Comparator | |
| AURELIA | Pujade-Lauraine (2014) | 01/01/2013 | INV | 182 | 179 | 166 | 135 | 0.48 (0.38, 0.60) |
| mEOC/GOG241‡ | Gore (2019) | 01/02/2018 | NR | 13 | 11 | NR | NR | 1.12 (0.45, 2.80) |
| | | | | 13 | 13 | NR | NR | 0.55 (0.21, 1.45) |
| *Cervical cancer* | | | | | | | | |
| GOG240 | Tewari (2014) | 01/03/2012 | NR | 225 | 227 | 184 | 183 | 0.67 (0.54, 0.82) |
| | Tewari (2017) | 01/03/2014 | NR | 225 | 227 | 206 | 199 | 0.68 (0.56, 0.84) |
| *Glioblastoma* | | | | | | | | |
| RTOG0825 | Gilbert (2014) | 01/12/2012 | NR | 317 | 320 | 256 | 256 | 0.79 (0.66, 0.94) |
| AvaGlio | Chinot (2014) | 01/03/2012 | IRC | 463 | 458 | 387 | 354 | 0.64 (0.55, 0.74) |
| EORTC26101 | Wick (2017) | 01/10/2015 | IRC | 149 | 288 | 143 | 260 | 0.49 (0.39, 0.61) |

† Where studies only reported month and year for the data cut-off, we assumed that this was the first of the month. ‡ Where studies reported multiple two-arm (chemotherapy vs. chemotherapy +bevacizumab) comparisons, both were included as long as there was no overlap in patients.

**Abbreviations:** AM, assessment method; CI, confidence interval; IRC, independent review committee; IRF, independent review facility; INV, investigator assessment; NR, not reported, NSCLC, non-small cell lung cancer.

# B: Statistical Methods and Results

## B-I: Description of statistical methods

The random-effects meta-analysis normal-normal hierarchical model[65] is used for within-indication meta-analysis. The relative treatment effect (for example the ln(HR)), $Y_{ij}$, is as assumed to follow a normal distribution:

$$Y_{ij} \sim N(\delta_{ij}, \sigma_{ij}^2) \qquad (1)$$

where $\delta_{ij}$ is the mean treatment effect and $\sigma_{ij}^2$ is the associated standard error for study $i$ within indication $j$. The mean treatment effect, $\delta_{ij}$ is assumed to be exchangeable across studies within each indication:

$$\delta_{ij} \sim N(d_j, \tau_j^2) \qquad (2)$$

where $d_j$ is the pooled treatment effect and $\tau_j$ is the between-study standard deviation, within-indication (heterogeneity). A weakly-informative half-normal prior distribution is place on the between-study standard deviation for each indication:[66]

$$\tau_j \sim |N(0, 0.5^2)| \qquad (3)$$

Assumptions on the degree of information sharing across indications differed for the three models we explored here:

1) <u>Independent parameter (IP) model</u>
   As there is no evidence sharing across indications, a vague normal prior distribution, $d_j \sim N(0, 1000)$ is used for the pooled, indication-specific relative treatment effect, $d_j$ for each indication.

2) <u>Common parameter (CP) model</u>
   In this model there is complete sharing of information, $d_j$ is replaced by a common parameter, $d$ in equation (2), which pools treatment effects across all indications. This common/pooled RTE is assigned a vague normal prior distribution, $d \sim N(0, 1000)$.

3) <u>Hierarchical meta-analysis (HMA) model</u>
   In the HMA model, we assume that indication-level parameters are fully exchangeable and vary according to a normal distribution: $N(m_d, \tau_d^2)$, where $m_d$ is the overall pooled effect and $\tau_d$ is the between-indication standard deviation. The pooled parameter $m_d$ is assigned a vague normal prior distribution and a weakly informative half-normal prior distribution is assigned to the standard deviation, $\tau_d$.

$$\begin{aligned} m_d &\sim N(0,1000) \\ \tau_d &\sim |N(0,0.5^2)| \end{aligned} \quad (4)$$

## B-II Results of Synthesis Models

**Table S4**. Synthesis results for overall survival. *Note: The treatment effect estimate is reported as the HR and corresponding 95% credible interval on the log-scale.*

| Time Point | | CP Model | IP Model | HMA Model |
|---|---|---|---|---|
| *Colorectal Cancer* | | | | |
| 31/12/2003<br>2 datapoints (2 in colorectal cancer) | Treatment Effect Estimate | -0.341 (-0.906, 0.253) | -0.341 (-0.906, 0.253) | -0.341 (-0.911, 0.260) |
| | Within-Indication SD | 0.207 (0.009, 0.889) | 0.207 (0.009, 0.889) | 0.210 (0.009, 0.898) |
| | Between-Indication SD | - | - | 1.003 (0.048, 1.950) |
| 31/12/2004<br>3 datapoints (3 in colorectal cancer) | Treatment Effect Estimate | -0.318 (-0.630, -0.002) | -0.318 (-0.630, -0.002) | -0.319 (-0.626, -0.008) |
| | Within-Indication SD | 0.117 (0.006, 0.627) | 0.117 (0.006, 0.627) | 0.116 (0.005, 0.629) |
| | Between-Indication SD | - | - | 0.997 (0.053, 1.949) |
| 31/12/2007<br>7 datapoints (4 in colorectal cancer) | Treatment Effect Estimate | -0.198 (-0.348, -0.041) | -0.248 (-0.502, -0.019) | -0.230 (-0.435, -0.030) |
| | Within-Indication SD | 0.120 (0.007, 0.463) | 0.130 (0.009, 0.527) | 0.124 (0.007, 0.488) |
| | Between-Indication SD | - | - | 0.202 (0.008, 1.544) |
| 31/12/2009<br>16 datapoints (6 in colorectal cancer) | Treatment Effect Estimate | -0.114 (-0.208, -0.019) | -0.171 (-0.374, 0.010) | -0.141 (-0.298, -0.004) |
| | Within-Indication SD | 0.151 (0.029, 0.392) | 0.161 (0.039, 0.435) | 0.154 (0.037, 0.403) |
| | Between-Indication SD | - | - | 0.089 (0.004, 0.675) |
| 31/12/2010<br>17 datapoints (7 in colorectal cancer) | Treatment Effect Estimate | -0.123 (-0.221, -0.026) | -0.196 (-0.392, -0.030) | -0.159 (-0.320, -0.026) |
| | Within-Indication SD | 0.157 (0.039, 0.391) | 0.164 (0.048, 0.415) | 0.158 (0.042, 0.392) |
| | Between-Indication SD | - | - | 0.097 (0.004, 0.689) |
| 31/12/2011<br>18 datapoints (8 in colorectal cancer) | Treatment Effect Estimate | -0.131 (-0.219, -0.041) | -0.190 (-0.350, -0.056) | -0.162 (-0.297, -0.048) |
| | Within-Indication SD | 0.133 (0.026, 0.330) | 0.138 (0.031, 0.340) | 0.133 (0.022, 0.330) |
| | Between-Indication SD | - | - | 0.096 (0.004, 0.687) |
| 31/12/2012<br>20 datapoints (9 in colorectal cancer) | Treatment Effect Estimate | -0.128 (-0.213, -0.038) | -0.191 (-0.332, -0.073) | -0.164 (-0.288, -0.054) |
| | Within-Indication SD | 0.125 (0.021, 0.307) | 0.127 (0.024, 0.300) | 0.123 (0.022, 0.295) |
| | Between-Indication SD | - | - | 0.099 (0.005, 0.519) |
| *Renal Cell Carcinoma* | | | | |

| Time Point | | CP Model | IP Model | HMA Model |
|---|---|---|---|---|
| 31/12/2008<br>10 datapoints (1 in renal cell carcinoma) | Treatment Effect Estimate | -0.166 (-0.285, -0.042) | -0.150 (-1.251, 0.939) | -0.160 (-0.494, 0.185) |
| | Within-Indication SD | 0.161 (0.007, 0.840) | 0.337 (0.016, 1.121) | 0.202 (0.008, 0.907) |
| | Between-Indication SD | - | - | 0.110 (0.005, 0.828) |
| 31/12/2009<br>16 datapoints (2 in renal cell carcinoma) | Treatment Effect Estimate | -0.114 (-0.208, -0.018) | -0.151 (-0.643, 0.340) | -0.126 (-0.333, 0.073) |
| | Within-Indication SD | 0.094 (0.004, 0.574) | 0.152 (0.006, 0.823) | 0.110 (0.005, 0.641) |
| | Between-Indication SD | - | - | 0.089 (0.004, 0.648) |
| *Breast Cancer* | | | | |
| 31/12/2007<br>7 datapoints (1 in breast cancer) | Treatment Effect Estimate | -0.197 (-0.348, -0.041) | -0.128 (-1.235, 0.992) | -0.166 (-0.664, 0.362) |
| | Within-Indication SD | 0.183 (0.008, 0.862) | 0.339 (0.016, 1.133) | 0.236 (0.010, 0.958) |
| | Between-Indication SD | - | - | 0.203 (0.008, 1.539) |
| 31/12/2008<br>11 datapoints (3 in breast cancer) | Treatment Effect Estimate | -0.166 (-0.285, -0.043) | -0.095 (-0.411, 0.230) | -0.133 (-0.322, 0.071) |
| | Within-Indication SD | 0.102 (0.005, 0.510) | 0.122 (0.006, 0.646) | 0.104 (0.004, 0.533) |
| | Between-Indication SD | - | - | 0.109 (0.004, 0.834) |
| 31/12/2009<br>16 datapoints (6 in breast cancer) | Treatment Effect Estimate | -0.113 (-0.209, -0.015) | -0.019 (-0.208, 0.240) | -0.071 (-0.210, 0.114) |
| | Within-Indication SD | 0.133 (0.006, 0.484) | 0.152 (0.009, 0.519) | 0.134 (0.007, 0.481) |
| | Between-Indication SD | - | - | 0.091 (0.004, 0.678) |
| 31/12/2012<br>20 datapoints (6 in breast cancer) | Treatment Effect Estimate | -0.128 (-0.215, -0.040) | -0.018 (-0.211, 0.242) | -0.071 (-0.212, 0.118) |
| | Within-Indication SD | 0.142 (0.007, 0.502) | 0.152 (0.008, 0.518) | 0.136 (0.007, 0.481) |
| | Between-Indication SD | - | - | 0.098 (0.005, 0.523) |
| 31/12/2013<br>26 datapoints (7 in breast cancer) | Treatment Effect Estimate | -0.115 (-0.188, -0.043) | -0.035 (-0.195, 0.173) | -0.085 (-0.189, 0.055) |
| | Within-Indication SD | 0.106 (0.005, 0.402) | 0.123 (0.006, 0.435) | 0.105 (0.004, 0.396) |
| | Between-Indication SD | - | - | 0.062 (0.003, 0.273) |
| 31/12/2015<br>32 datapoints (10 in breast cancer) | Treatment Effect Estimate | -0.113 (-0.174, -0.054) | -0.055 (-0.168, 0.076) | -0.091 (-0.174, 0.010) |
| | Within-Indication SD | 0.068 (0.003, 0.270) | 0.073 (0.003, 0.280) | 0.066 (0.003, 0.263) |
| | Between-Indication SD | - | - | 0.052 (0.003, 0.208) |
| *Non-small cell lung cancer* | | | | |

| Time Point | | CP Model | IP Model | HMA Model |
|---|---|---|---|---|
| 31/12/2005<br>4 datapoints (1 in NSCLC) | Treatment Effect Estimate | -0.296 (-0.501, -0.084) | -0.235 (-1.335, 0.856) | -0.259 (-0.972, 0.450) |
| | Within-Indication SD | 0.188 (0.008, 0.872) | 0.335 (0.015, 1.120) | 0.271 (0.011, 1.015) |
| | Between-Indication SD | - | - | 0.475 (0.016, 1.872) |
| 31/12/2007<br>7 datapoints (2 in NSCLC) | Treatment Effect Estimate | -0.197 (-0.348, -0.041) | -0.108 (-0.731, 0.520) | -0.150 (-0.507, 0.237) |
| | Within-Indication SD | 0.208 (0.020, 0.758) | 0.269 (0.028, 0.945) | 0.230 (0.023, 0.826) |
| | Between-Indication SD | - | - | 0.201 (0.009, 1.542) |
| 31/12/2013<br>26 datapoints (3 in NSCLC) | Treatment Effect Estimate | -0.116 (-0.187, -0.043) | -0.174 (-0.625, 0.243) | -0.125 (-0.296, 0.027) |
| | Within-Indication SD | 0.173 (0.016, 0.609) | 0.235 (0.028, 0.796) | 0.186 (0.017, 0.643) |
| | Between-Indication SD | - | - | 0.062 (0.003, 0.271) |
| 31/12/2017<br>33 datapoints (4 in NSCLC) | Treatment Effect Estimate | -0.116 (-0.176, -0.056) | -0.177 (-0.501, 0.114) | -0.130 (-0.275, -0.014) |
| | Within-Indication SD | 0.148 (0.012, 0.497) | 0.185 (0.018, 0.626) | 0.153 (0.013, 0.511) |
| | Between-Indication SD | - | - | 0.052 (0.003, 0.208) |
| 31/12/2018<br>37 datapoints (5 in NSCLC) | Treatment Effect Estimate | -0.104 (-0.161, -0.048) | -0.156 (-0.412, 0.103) | -0.118 (-0.252, -0.009) |
| | Within-Indication SD | 0.139 (0.011, 0.446) | 0.168 (0.016, 0.540) | 0.145 (0.013, 0.456) |
| | Between-Indication SD | - | - | 0.050 (0.002, 0.200) |
| 31/12/2019<br>38 datapoints (6 in NSCLC) | Treatment Effect Estimate | -0.104 (-0.161, -0.048) | -0.137 (-0.346, 0.080) | -0.114 (-0.232, -0.011) |
| | Within-Indication SD | 0.124 (0.008, 0.381) | 0.149 (0.014, 0.456) | 0.130 (0.010, 0.391) |
| | Between-Indication SD | - | - | 0.049 (0.003, 0.194) |
| *Ovarian, Fallopian Tube and Primary Peritoneal Cancer* | | | | |
| 31/12/2013<br>26 datapoints (3 in OFTPP cancer) | Treatment Effect Estimate | -0.115 (-0.187, -0.042) | -0.058 (-0.353, 0.210) | -0.092 (-0.219, 0.042) |
| | Within-Indication SD | 0.088 (0.004, 0.427) | 0.101 (0.004, 0.598) | 0.087 (0.004, 0.448) |
| | Between-Indication SD | - | - | 0.063 (0.003, 0.267) |
| 31/12/2014<br>29 datapoints (4 in OFTPP cancer) | Treatment Effect Estimate | -0.124 (-0.189, -0.060) | -0.090 (-0.282, 0.087) | -0.113 (-0.214, -0.005) |
| | Within-Indication SD | 0.075 (0.004, 0.321) | 0.084 (0.003, 0.418) | 0.075 (0.004, 0.336) |
| | Between-Indication SD | - | - | 0.050 (0.003, 0.214) |
| 31/12/2018 | Treatment Effect Estimate | -0.105 (-0.161, -0.050) | -0.070 (-0.190, 0.042) | -0.092 (-0.173, -0.006) |

| Time Point | | CP Model | IP Model | HMA Model |
|---|---|---|---|---|
| 37 datapoints (7 in OFTPP cancer) | Within-Indication SD | 0.056 (0.003, 0.221) | 0.056 (0.003, 0.255) | 0.053 (0.003, 0.225) |
| | Between-Indication SD | - | - | 0.049 (0.002, 0.200) |
| *Cervical Cancer* | | | | |
| 31/12/2014 29 datapoints (1 in cervical cancer) | Treatment Effect Estimate | -0.125 (-0.189, -0.059) | -0.262 (-1.367, 0.842) | -0.133 (-0.330, 0.023) |
| | Within-Indication SD | 0.214 (0.010, 0.899) | 0.334 (0.016, 1.120) | 0.215 (0.010, 0.891) |
| | Between-Indication SD | - | - | 0.050 (0.003, 0.216) |
| *Glioblastoma* | | | | |
| 31/12/2012 20 datapoints (1 in glioblastoma) | Treatment Effect Estimate | -0.128 (-0.213, -0.040) | 0.121 (-0.991, 1.217) | -0.083 (-0.345, 0.285) |
| | Within-Indication SD | 0.320 (0.043, 0.977) | 0.338 (0.016, 1.119) | 0.286 (0.018, 0.972) |
| | Between-Indication SD | - | - | 0.099 (0.005, 0.517) |
| 31/12/2013 26 datapoints (2 in glioblastoma) | Treatment Effect Estimate | -0.115 (-0.188, -0.042) | -0.010 (-0.615, 0.609) | -0.094 (-0.256, 0.097) |
| | Within-Indication SD | 0.192 (0.015, 0.729) | 0.257 (0.023, 0.931) | 0.194 (0.014, 0.746) |
| | Between-Indication SD | - | - | 0.064 (0.003, 0.272) |
| 31/12/2015 32 datapoints (3 in glioblastoma) | Treatment Effect Estimate | -0.114 (-0.174, -0.054) | -0.028 (-0.365, 0.323) | -0.097 (-0.217, 0.051) |
| | Within-Indication SD | 0.139 (0.009, 0.544) | 0.160 (0.011, 0.682) | 0.138 (0.008, 0.551) |
| | Between-Indication SD | - | - | 0.052 (0.003, 0.212) |

All reported estimates are the median and the corresponding 95% credible interval

**Abbreviations:** CP, common parameter; HMA, hierarchical meta-analysis; HR, hazard ratio; IP, independent parameter; OFTPP, ovarian, fallopian tube and primary peritoneal cancer; NSCLC; non-small cell lung cancer; SD, standard deviation.

**Table S5**. Synthesis results for progression-free survival. *Note: The treatment effect estimate is reported as the HR and corresponding 95% credible interval on the log-scale.*

| Time Point | | CP Model | IP Model | HMA Model |
|---|---|---|---|---|
| *Colorectal Cancer* | | | | |
| 31/12/2003<br>5 datapoints (3 in colorectal cancer) | Treatment Effect Estimate | -0.653 (-0.975, -0.254) | -0.684 (-1.082, -0.289) | -0.666 (-1.034, -0.279) |
| | Within-Indication SD | 0.142 (0.006, 0.666) | 0.145 (0.006, 0.706) | 0.146 (0.006, 0.682) |
| | Between-Indication SD | - | - | 0.552 (0.032, 1.823) |
| 31/12/2004<br>6 datapoints (4 in colorectal cancer) | Treatment Effect Estimate | -0.592 (-0.850, -0.324) | -0.604 (-0.907, -0.343) | -0.597 (-0.880, -0.331) |
| | Within-Indication SD | 0.116 (0.005, 0.530) | 0.121 (0.005, 0.550) | 0.119 (0.005, 0.540) |
| | Between-Indication SD | - | - | 0.534 (0.031, 1.817) |
| 31/12/2006<br>9 datapoints (5 in colorectal cancer) | Treatment Effect Estimate | -0.463 (-0.662, -0.278) | -0.488 (-0.845, -0.200) | -0.470 (-0.755, -0.232) |
| | Within-Indication SD | 0.224 (0.070, 0.574) | 0.245 (0.075, 0.641) | 0.234 (0.071, 0.606) |
| | Between-Indication SD | - | - | 0.218 (0.009, 1.299) |
| 31/12/2009<br>20 datapoints (7 in colorectal cancer) | Treatment Effect Estimate | -0.357 (-0.489, -0.227) | -0.415 (-0.691, -0.190) | -0.386 (-0.604, -0.212) |
| | Within-Indication SD | 0.225 (0.096, 0.501) | 0.246 (0.109, 0.554) | 0.233 (0.102, 0.520) |
| | Between-Indication SD | - | - | 0.140 (0.007, 0.900) |
| 31/12/2010<br>21 datapoints (8 in colorectal cancer) | Treatment Effect Estimate | -0.380 (-0.512, -0.244) | -0.467 (-0.730, -0.245) | -0.424 (-0.648, -0.244) |
| | Within-Indication SD | 0.249 (0.120, 0.516) | 0.264 (0.132, 0.544) | 0.255 (0.127, 0.523) |
| | Between-Indication SD | - | - | 0.156 (0.009, 0.954) |
| 31/12/2011<br>26 datapoints (9 in colorectal cancer) | Treatment Effect Estimate | -0.393 (-0.510, -0.273) | -0.450 (-0.676, -0.265) | -0.424 (-0.609, -0.270) |
| | Within-Indication SD | 0.221 (0.109, 0.448) | 0.237 (0.118, 0.479) | 0.228 (0.113, 0.458) |
| | Between-Indication SD | - | - | 0.138 (0.009, 0.625) |
| 31/12/2012<br>29 datapoints (10 in colorectal cancer) | Treatment Effect Estimate | -0.393 (-0.497, -0.290) | -0.469 (-0.671, -0.297) | -0.430 (-0.600, -0.294) |
| | Within-Indication SD | 0.218 (0.112, 0.427) | 0.230 (0.122, 0.443) | 0.222 (0.117, 0.428) |
| | Between-Indication SD | - | - | 0.108 (0.006, 0.429) |
| *Renal Cell Carcinoma* | | | | |
| 31/12/2003<br>5 datapoints (1 in renal cell carcinoma) | Treatment Effect Estimate | -0.653 (-0.973, -0.256) | -0.935 (-2.123, 0.261) | -0.757 (-1.616, 0.060) |
| | Within-Indication SD | 0.287 (0.013, 0.997) | 0.337 (0.016, 1.119) | 0.312 (0.014, 1.054) |

| Time Point | | CP Model | IP Model | HMA Model |
|---|---|---|---|---|
| | Between-Indication SD | - | - | 0.553 (0.030, 1.829) |
| 31/12/2008<br>14 datapoints (3 in renal cell carcinoma) | Treatment Effect Estimate | -0.408 (-0.546, -0.282) | -0.503 (-1.001, -0.152) | -0.447 (-0.718, -0.239) |
| | Within-Indication SD | 0.133 (0.005, 0.641) | 0.186 (0.008, 0.778) | 0.146 (0.005, 0.674) |
| | Between-Indication SD | - | - | 0.115 (0.005, 0.832) |
| 31/12/2009<br>20 datapoints (3 in renal cell carcinoma) | Treatment Effect Estimate | -0.357 (-0.490, -0.227) | -0.504 (-1.012, -0.147) | -0.429 (-0.710, -0.195) |
| | Within-Indication SD | 0.179 (0.008, 0.712) | 0.188 (0.008, 0.789) | 0.159 (0.007, 0.701) |
| | Between-Indication SD | - | - | 0.143 (0.008, 0.895) |
| *Breast Cancer* | | | | |
| 31/12/2002<br>2 datapoints (1 in breast cancer) | Treatment Effect Estimate | -0.300 (-1.135, 0.317) | -0.018 (-1.130, 1.086) | -0.099 (-1.142, 0.723) |
| | Within-Indication SD | 0.387 (0.020, 1.110) | 0.338 (0.015, 1.123) | 0.334 (0.016, 1.098) |
| | Between-Indication SD | - | - | 0.799 (0.043, 1.918) |
| 31/12/2007<br>12 datapoints (2 in breast cancer) | Treatment Effect Estimate | -0.423 (-0.582, -0.271) | -0.249 (-1.014, 0.479) | -0.373 (-0.708, 0.030) |
| | Within-Indication SD | 0.331 (0.074, 0.902) | 0.366 (0.052, 1.032) | 0.329 (0.056, 0.917) |
| | Between-Indication SD | - | - | 0.138 (0.006, 0.957) |
| 31/12/2008<br>14 datapoints (4 in breast cancer) | Treatment Effect Estimate | -0.408 (-0.548, -0.281) | -0.333 (-0.667, -0.005) | -0.376 (-0.590, -0.154) |
| | Within-Indication SD | 0.199 (0.023, 0.586) | 0.216 (0.026, 0.670) | 0.201 (0.023, 0.601) |
| | Between-Indication SD | - | - | 0.114 (0.005, 0.803) |
| 31/12/2009<br>20 datapoints (8 in breast cancer) | Treatment Effect Estimate | -0.357 (-0.491, -0.226) | -0.212 (-0.459, 0.045) | -0.281 (-0.482, -0.051) |
| | Within-Indication SD | 0.308 (0.144, 0.611) | 0.288 (0.132, 0.591) | 0.289 (0.132, 0.585) |
| | Between-Indication SD | - | - | 0.143 (0.007, 0.919) |
| 31/12/2011<br>26 datapoints (9 in breast cancer) | Treatment Effect Estimate | -0.393 (-0.511, -0.273) | -0.212 (-0.421, 0.008) | -0.287 (-0.483, -0.085) |
| | Within-Indication SD | 0.310 (0.149, 0.591) | 0.257 (0.113, 0.521) | 0.266 (0.115, 0.536) |
| | Between-Indication SD | - | - | 0.135 (0.009, 0.606) |
| 31/12/2013<br>33 datapoints (11 in breast cancer) | Treatment Effect Estimate | -0.387 (-0.485, -0.285) | -0.220 (-0.377, -0.053) | -0.284 (-0.448, -0.127) |
| | Within-Indication SD | 0.272 (0.128, 0.506) | 0.210 (0.075, 0.418) | 0.220 (0.076, 0.441) |
| | Between-Indication SD | - | - | 0.110 (0.008, 0.408) |

| Time Point | | CP Model | IP Model | HMA Model |
|---|---|---|---|---|
| 31/12/2014<br>37 datapoints (12 in breast cancer) | Treatment Effect Estimate | -0.390 (-0.479, -0.299) | -0.233 (-0.377, -0.081) | -0.300 (-0.453, -0.153) |
| | Within-Indication SD | 0.257 (0.120, 0.469) | 0.198 (0.068, 0.390) | 0.212 (0.075, 0.417) |
| | Between-Indication SD | - | - | 0.094 (0.006, 0.320) |
| 31/12/2015<br>39 datapoints (13 in breast cancer) | Treatment Effect Estimate | -0.402 (-0.490, -0.309) | -0.236 (-0.373, -0.094) | -0.300 (-0.455, -0.160) |
| | Within-Indication SD | 0.258 (0.125, 0.462) | 0.189 (0.061, 0.367) | 0.203 (0.071, 0.400) |
| | Between-Indication SD | - | - | 0.098 (0.007, 0.326) |
| *Non-small cell lung cancer* | | | | |
| 31/12/2005<br>8 datapoints (1 in NSCLC) | Treatment Effect Estimate | -0.551 (-0.774, -0.355) | -0.415 (-1.513, 0.686) | -0.492 (-1.014, 0.014) |
| | Within-Indication SD | 0.238 (0.012, 0.922) | 0.339 (0.016, 1.124) | 0.257 (0.012, 0.973) |
| | Between-Indication SD | - | - | 0.222 (0.010, 1.341) |
| 31/12/2007<br>13 datapoints (2 in NSCLC) | Treatment Effect Estimate | -0.423 (-0.572, -0.284) | -0.291 (-0.913, 0.329) | -0.373 (-0.663, -0.066) |
| | Within-Indication SD | 0.241 (0.037, 0.800) | 0.265 (0.030, 0.936) | 0.233 (0.033, 0.813) |
| | Between-Indication SD | - | - | 0.121 (0.005, 0.881) |
| 31/12/2013<br>35 datapoints (4 in NSCLC) | Treatment Effect Estimate | -0.392 (-0.485, -0.298) | -0.496 (-0.969, -0.067) | -0.415 (-0.656, -0.225) |
| | Within-Indication SD | 0.298 (0.108, 0.728) | 0.341 (0.134, 0.835) | 0.307 (0.117, 0.738) |
| | Between-Indication SD | - | - | 0.096 (0.006, 0.332) |
| 31/12/2014<br>37 datapoints (4 in NSCLC) | Treatment Effect Estimate | -0.390 (-0.479, -0.299) | -0.504 (-0.976, -0.073) | -0.416 (-0.657, -0.228) |
| | Within-Indication SD | 0.304 (0.113, 0.733) | 0.343 (0.136, 0.835) | 0.311 (0.120, 0.749) |
| | Between-Indication SD | - | - | 0.094 (0.006, 0.317) |
| 31/12/2018<br>43 datapoints (6 in NSCLC) | Treatment Effect Estimate | -0.399 (-0.483, -0.310) | -0.451 (-0.768, -0.152) | -0.415 (-0.605, -0.252) |
| | Within-Indication SD | 0.255 (0.096, 0.580) | 0.284 (0.113, 0.653) | 0.264 (0.104, 0.599) |
| | Between-Indication SD | - | - | 0.090 (0.005, 0.299) |
| 31/12/2019<br>43 datapoints (6 in NSCLC) | Treatment Effect Estimate | -0.394 (-0.478, -0.305) | -0.413 (-0.732, -0.107) | -0.400 (-0.585, -0.233) |
| | Within-Indication SD | 0.256 (0.097, 0.582) | 0.290 (0.113, 0.670) | 0.267 (0.103, 0.598) |
| | Between-Indication SD | - | - | 0.088 (0.006, 0.296) |
| *Ovarian, Fallopian Tube and Primary Peritoneal Cancer* | | | | |

| Time Point | | CP Model | IP Model | HMA Model |
|---|---|---|---|---|
| 31/12/2011<br>26 datapoints (3 in OFTPP) | Treatment Effect Estimate | -0.393 (-0.510, -0.273) | -0.558 (-1.091, -0.047) | -0.443 (-0.764, -0.186) |
| | Within-Indication SD | 0.313 (0.119, 0.795) | 0.330 (0.125, 0.887) | 0.310 (0.119, 0.804) |
| | Between-Indication SD | - | - | 0.137 (0.007, 0.624) |
| 31/12/2013<br>33 datapoints (4 in OFTPP) | Treatment Effect Estimate | -0.387 (-0.486, -0.286) | -0.433 (-0.908, 0.020) | -0.397 (-0.649, -0.177) |
| | Within-Indication SD | 0.323 (0.158, 0.739) | 0.369 (0.174, 0.859) | 0.337 (0.163, 0.768) |
| | Between-Indication SD | - | - | 0.111 (0.007, 0.415) |
| 31/12/2014<br>37 datapoints (5 in OFTPP) | Treatment Effect Estimate | -0.390 (-0.479, -0.300) | -0.437 (-0.800, -0.090) | -0.403 (-0.610, -0.226) |
| | Within-Indication SD | 0.283 (0.144, 0.630) | 0.318 (0.157, 0.728) | 0.294 (0.148, 0.654) |
| | Between-Indication SD | - | - | 0.094 (0.006, 0.322) |
| 31/12/2018<br>42 datapoints (7 in OFTPP) | Treatment Effect Estimate | -0.397 (-0.483, -0.305) | -0.415 (-0.712, -0.113) | -0.403 (-0.588, -0.232) |
| | Within-Indication SD | 0.272 (0.141, 0.573) | 0.300 (0.152, 0.646) | 0.281 (0.145, 0.589) |
| | Between-Indication SD | - | - | 0.093 (0.008, 0.310) |
| *Cervical Cancer* | | | | |
| 31/12/2014<br>37 datapoints (1 in cervical cancer) | Treatment Effect Estimate | -0.389 (-0.480, -0.300) | -0.385 (-1.485, 0.702) | -0.391 (-0.636, -0.159) |
| | Within-Indication SD | 0.161 (0.007, 0.832) | 0.337 (0.015, 1.114) | 0.189 (0.008, 0.876) |
| | Between-Indication SD | - | - | 0.093 (0.006, 0.325) |
| *Glioblastoma* | | | | |
| 31/12/2012<br>29 datapoints (2 in glioblastoma) | Treatment Effect Estimate | -0.394 (-0.497, -0.291) | -0.348 (-0.926, 0.248) | -0.375 (-0.592, -0.152) |
| | Within-Indication SD | 0.150 (0.009, 0.670) | 0.231 (0.015, 0.910 | 0.166 (0.010, 0.715) |
| | Between-Indication SD | - | - | 0.108 (0.007, 0.422) |
| 31/12/2015<br>39 datapoints (3 in glioblastoma) | Treatment Effect Estimate | -0.403 (-0.490, -0.309) | -0.455 (-0.930, 0.011) | -0.420 (-0.634, -0.229) |
| | Within-Indication SD | 0.217 (0.040, 0.665) | 0.277 (0.065, 0.828) | 0.234 (0.047, 0.694) |
| | Between-Indication SD | - | - | 0.099 (0.008, 0.323) |

All reported estimates are the median and corresponding 95% credible interval

**Abbreviations:** CP, common parameter; HMA, hierarchical meta-analysis; HR, hazard ratio; IP, independent parameter; OFTPP, ovarian, fallopian tube and primary peritoneal cancer; NSCLC; non-small cell lung cancer; SD, standard deviation

# Model Fit Statistics

Model fit was assessed using the DIC and total residual deviance. Model fit statistics for the analyses conducted in Tables S4 and S5 are reported in Table S6 and Table S7, respectively.

All three models fit the data reasonably well; the total residual deviance consistent with the number of datapoints included in the analysis except for earlier timepoints in the PFS analyses, likely due to the sparsity of evidence.

The DICs were consistent across the three models, suggesting that all models were appropriate and comparable.

**Table S6**. Model fit statistics for the overall survival analyses

| Time Point | | CP Model | IP Model | HMA Model |
|---|---|---|---|---|
| *Colorectal Cancer* | | | | |
| 31/12/2003 (2 datapoints) | DIC | 0.7704 | 0.7704 | 0.7645 |
|  | pD | 1.647 | 1.647 | 1.647 |
|  | Deviance | -0.877 | -0.877 | -0.883 |
| 31/12/2004 (3 datapoints) | DIC | -1.354 | -1.354 | -1.328 |
|  | pD | 2.016 | 2.016 | 2.026 |
|  | Deviance | -3.370 | -3.370 | -3.354 |
| 31/12/2007 (7 datapoints) | DIC | -2.595 | -2.188 | -2.37 |
|  | pD | 5.562 | 6.053 | 5.862 |
|  | Deviance | -8.157 | -8.241 | -8.231 |
| 31/12/2009 (16 datapoints) | DIC | -4.478 | -3.861 | -4.353 |
|  | pD | 11.08 | 12.58 | 11.73 |
|  | Deviance | -15.558 | -16.441 | -16.088 |
| 31/12/2010 (17 datapoints) | DIC | -3.838 | -3.611 | -3.898 |
|  | pD | 11.69 | 13.06 | 12.26 |
|  | Deviance | -15.526 | -16.672 | -16.156 |
| 31/12/2011 (18 datapoints) | DIC | -5.385 | -5.289 | -5.445 |
|  | pD | 12.08 | 13.43 | 12.65 |
|  | Deviance | -17.461 | -18.722 | -18.099 |
| 31/12/2012 (20 datapoints) | DIC | -5.239 | -5.61 | -5.706 |
|  | pD | 13.32 | 14.63 | 13.79 |
|  | Deviance | -18.561 | -20.243 | -19.492 |
| *Renal Cell Carcinoma* | | | | |
| 31/12/2008 (10 datapoints) | DIC | -4.182 | -3.248 | -3.873 |
|  | pD | 7.066 | 8.067 | 7.515 |
|  | Deviance | -11.248 | -11.315 | -11.388 |
| 31/12/2009 (16 datapoints) | DIC | -4.495 | -3.845 | -4.281 |
|  | pD | 11.09 | 12.58 | 11.74 |
|  | Deviance | -15.586 | -16.429 | -16.017 |
| *Breast Cancer* | | | | |
| 31/12/2007 (7 datapoints) | DIC | -2.577 | -2.161 | -2.320 |
|  | pD | 5.57 | 6.06 | 5.89 |

| Time Point | | CP Model | IP Model | HMA Model |
|---|---|---|---|---|
| | Deviance | -8.147 | -8.224 | -8.210 |
| 31/12/2008 | DIC | -4.141 | -3.196 | -3.840 |
| (11 datapoints) | pD | 7.079 | 8.098 | 7.525 |
| | Deviance | -11.220 | -11.294 | -11.365 |
| 31/12/2009 | DIC | -4.479 | -3.899 | -4.305 |
| (16 datapoints) | pD | 11.11 | 12.59 | 11.74 |
| | Deviance | -15.585 | -16.494 | -16.046 |
| 31/12/2012 | DIC | -5.204 | -5.563 | -5.809 |
| (20 datapoints) | pD | 13.32 | 14.68 | 13.81 |
| | Deviance | -18.525 | -20.239 | -19.623 |
| 31/12/2013 | DIC | -10.18 | -9.537 | -10.44 |
| (26 datapoints) | pD | 16.1 | 18.47 | 16.75 |
| | Deviance | -26.284 | -28.003 | -27.187 |
| 31/12/2015 | DIC | -15.69 | -14.04 | -15.64 |
| (32 datapoints) | pD | 17.12 | 20.11 | 17.96 |
| | Deviance | -32.817 | -34.151 | -33.599 |
| *Non-small cell lung cancer* | | | | |
| 31/12/2005 | DIC | -1.721 | -1.154 | -1.297 |
| (4 datapoints) | pD | 2.692 | 3.021 | 2.939 |
| | Deviance | -4.413 | -4.174 | -4.236 |
| 31/12/2007 | DIC | -2.596 | -2.193 | -2.352 |
| (7 datapoints) | pD | 5.565 | 6.046 | 5.862 |
| | Deviance | -8.161 | -8.239 | -8.215 |
| 31/12/2013 | DIC | -10.18 | -9.502 | -10.44 |
| (26 datapoints) | pD | 16.06 | 18.49 | 16.79 |
| | Deviance | -26.242 | -27.994 | -27.228 |
| 31/12/2017 | DIC | -15.97 | -14.11 | -15.97 |
| (33 datapoints) | pD | 17.35 | 20.58 | 18.28 |
| | Deviance | -33.319 | -34.686 | -34.259 |
| 31/12/2018 | DIC | -16.68 | -14.87 | -16.72 |
| (37 datapoints) | pD | 17.72 | 20.92 | 18.62 |
| | Deviance | -34.395 | -35.785 | -35.334 |
| 31/12/2019 | DIC | -17.27 | -15.28 | -17.14 |
| (38 datapoints) | pD | 17.85 | 21.13 | 18.81 |
| | Deviance | -35.120 | -36.413 | -35.950 |
| *Ovarian, Fallopian Tube and Primary Peritoneal Cancer* | | | | |
| 31/12/2013 | DIC | -10.21 | -9.538 | -10.4 |
| (26 datapoints) | pD | 16.11 | 18.43 | 16.77 |
| | Deviance | -26.316 | -27.963 | -27.175 |
| 31/12/2014 | DIC | -12.65 | -10.87 | -12.56 |
| (29 datapoints) | pD | 16.91 | 19.98 | 17.78 |

| Time Point | | CP Model | IP Model | HMA Model |
|---|---|---|---|---|
| | Deviance | -29.563 | -30.843 | -30.34 |
| 31/12/2018 | DIC | -16.62 | -14.81 | -16.66 |
| (37 datapoints) | pD | 17.72 | 20.91 | 18.63 |
| | Deviance | -34.340 | -35.712 | -35.295 |
| *Cervical Cancer* | | | | |
| 31/12/2014 | DIC | -12.61 | -10.86 | -12.5 |
| (29 datapoints) | pD | 17.02 | 19.97 | 17.82 |
| | Deviance | -29.628 | -30.831 | -30.318 |
| *Glioblastoma* | | | | |
| 31/12/2012 | DIC | -5.28 | -5.558 | -5.726 |
| (20 datapoints) | pD | 13.30 | 14.69 | 13.81 |
| | Deviance | -18.579 | -20.244 | -19.537 |
| 31/12/2013 | DIC | -10.200 | -9.432 | -10.410 |
| (26 datapoints) | pD | 16.13 | 18.46 | 16.85 |
| | Deviance | -26.336 | -27.894 | -27.253 |
| 31/12/2015 | DIC | -15.68 | -13.95 | -15.61 |
| (32 datapoints) | pD | 17.07 | 20.23 | 18.03 |
| | Deviance | -32.745 | -34.178 | -33.642 |

**Abbreviations:** CP, common parameter; HMA, hierarchical meta-analysis; IP, independent parameter; NSCLC; non-small cell lung cancer.

**Table S7**. Model fit statistics for the progression-free survival analyses

| Time Point | | CP Model | IP Model | HMA Model |
|---|---|---|---|---|
| *Colorectal Cancer* | | | | |
| 31/12/2003 (5 datapoints) | DIC | 4.659 | 5.038 | 4.913 |
| | pD | 3.342 | 3.734 | 3.621 |
| | Deviance | 1.317 | 1.305 | 1.291 |
| 31/12/2004 (6 datapoints) | DIC | 3.717 | 3.894 | 3.817 |
| | pD | 3.758 | 4.118 | 4.019 |
| | Deviance | -0.042 | -0.224 | -0.201 |
| 31/12/2006 (9 datapoints) | DIC | 3.273 | 3.552 | 3.444 |
| | pD | 7.066 | 7.758 | 7.428 |
| | Deviance | -3.793 | -4.207 | -3.985 |
| 31/12/2009 (20 datapoints) | DIC | -1.918 | -2.69 | -2.424 |
| | pD | 16.67 | 17.28 | 16.88 |
| | Deviance | -18.588 | -19.969 | -19.302 |
| 31/12/2010 (21 datapoints) | DIC | -1.906 | -2.74 | -2.417 |
| | pD | 17.47 | 18.04 | 17.70 |
| | Deviance | -19.38 | -20.782 | -20.118 |
| 31/12/2011 (26 datapoints) | DIC | -5.401 | -5.998 | -5.725 |
| | pD | 21.68 | 22.34 | 21.91 |
| | Deviance | -27.085 | -28.333 | -27.64 |
| 31/12/2012 (29 datapoints) | DIC | -7.013 | -7.575 | -7.345 |
| | pD | 24.00 | 24.98 | 24.34 |
| | Deviance | -31.015 | -32.555 | -31.681 |
| *Renal Cell Carcinoma* | | | | |
| 31/12/2003 (5 datapoints) | DIC | 4.672 | 5.078 | 4.892 |
| | pD | 3.349 | 3.756 | 3.615 |
| | Deviance | 1.323 | 1.322 | 1.277 |
| 31/12/2008 (14 datapoints) | DIC | 0.2667 | 0.4393 | 0.2947 |
| | pD | 10.91 | 12.06 | 11.40 |
| | Deviance | -10.646 | -11.621 | -11.102 |
| 31/12/2009 (20 datapoints) | DIC | -1.88 | -2.709 | -2.456 |
| | pD | 16.7 | 17.25 | 16.88 |
| | Deviance | -18.577 | -19.964 | -19.335 |
| *Breast Cancer* | | | | |
| 31/12/2002 (2 datapoints) | DIC | 3.903 | 3.682 | 3.67 |
| | pD | 1.948 | 2.002 | 1.978 |
| | Deviance | 1.956 | 1.679 | 1.692 |
| 31/12/2007 (12 datapoints) | DIC | 1.571 | 1.968 | 1.734 |
| | pD | 9.542 | 10.55 | 9.984 |
| | Deviance | -7.972 | -8.579 | -8.249 |
| 31/12/2008 | DIC | 0.3024 | 0.3527 | 0.2244 |
| (14 datapoints) | pD | 10.94 | 12.03 | 11.37 |
| | Deviance | -10.634 | -11.678 | -11.147 |
| 31/12/2009 (20 datapoints) | DIC | -1.929 | -2.704 | -2.426 |
| | pD | 16.69 | 17.28 | 16.91 |
| | Deviance | -18.619 | -19.985 | -19.334 |
| 31/12/2011 (26 datapoints) | DIC | -5.35 | -5.935 | -5.651 |
| | pD | 21.71 | 22.34 | 21.92 |
| | Deviance | -27.061 | -28.275 | -27.57 |
| 31/12/2013 (33 datapoints) | DIC | -9.362 | -10.05 | -9.588 |
| | pD | 27.45 | 28.03 | 27.52 |
| | Deviance | -36.811 | -38.086 | -37.103 |
| 31/12/2014 (37 datapoints) | DIC | -10.65 | -10.70 | -10.81 |
| | pD | 30.17 | 31.11 | 30.31 |
| | Deviance | -40.825 | -41.816 | -41.112 |
| 31/12/2015 (39 datapoints) | DIC | -10.66 | -10.74 | -10.82 |
| | pD | 31.72 | 32.32 | 31.64 |
| | Deviance | -42.376 | -43.065 | -42.463 |
| *Non-small cell lung cancer* | | | | |
| 31/12/2005 (8 datapoints) | DIC | 4.078 | 4.92 | 4.354 |
| | pD | 5.435 | 6.222 | 5.797 |
| | Deviance | -1.357 | -1.302 | -1.443 |
| 31/12/2007 (13 datapoints) | DIC | 1.09 | 1.498 | 1.207 |
| | pD | 10.29 | 11.39 | 10.73 |
| | Deviance | -9.196 | -9.887 | -9.52 |
| 31/12/2013 (35 datapoints) | DIC | -9.238 | -9.029 | -9.267 |
| | pD | 28.71 | 29.8 | 28.89 |
| | Deviance | -37.952 | -38.83 | -38.159 |
| 31/12/2014 (37 datapoints) | DIC | -10.69 | -10.54 | -10.72 |
| | pD | 30.16 | 31.16 | 30.30 |
| | Deviance | -40.848 | -41.693 | -41.017 |
| 31/12/2018 (43 datapoints) | DIC | -7.556 | -6.862 | -7.313 |
| | pD | 33.33 | 34.37 | 33.47 |
| | Deviance | -40.889 | -41.231 | -40.783 |
| 31/12/2019 (43 datapoints) | DIC | -7.415 | -7.059 | -7.443 |
| | pD | 33.44 | 34.45 | 33.51 |
| | Deviance | -40.855 | -41.510 | -40.958 |
| *Ovarian, Fallopian Tube and Primary Peritoneal Cancer* | | | | |
| 31/12/2011 (26 datapoints) | DIC | -5.382 | -6.008 | -5.678 |
| | pD | 21.68 | 22.33 | 21.92 |
| | Deviance | -27.066 | -28.336 | -27.600 |
| 31/12/2013 (33 datapoints) | DIC | -9.312 | -10.12 | -9.712 |
| | pD | 27.47 | 28.04 | 27.52 |
|  | Deviance | -36.778 | -38.161 | -37.233 |
| 31/12/2014<br>(37 datapoints) | DIC | -10.75 | -10.56 | -10.70 |
|  | pD | 30.15 | 31.13 | 30.32 |
|  | Deviance | -40.898 | -41.692 | -41.017 |
| 31/12/2018<br>(42 datapoints) | DIC | -7.346 | -7.043 | -7.32 |
|  | pD | 32.87 | 33.75 | 32.91 |
|  | Deviance | -40.216 | -40.796 | -40.228 |
| *Cervical Cancer* |  |  |  |  |
| 31/12/2014<br>(37 datapoints) | DIC | -10.70 | -10.64 | -10.79 |
|  | pD | 30.16 | 31.14 | 30.3 |
|  | Deviance | -40.853 | -41.784 | -41.087 |
| *Glioblastoma* |  |  |  |  |
| 31/12/2012<br>(29 datapoints) | DIC | -6.988 | -7.521 | -7.294 |
|  | pD | 24.01 | 24.98 | 24.35 |
|  | Deviance | -30.999 | -32.500 | -31.648 |
| 31/12/2015<br>(39 datapoints) | DIC | -10.75 | -10.70 | -10.75 |
|  | pD | 31.70 | 32.33 | 31.67 |
|  | Deviance | -42.447 | -43.027 | -42.417 |

**Abbreviations:** CP, common parameter; HMA, hierarchical meta-analysis; IP, independent parameter; NSCLC; non-small cell lung cancer.

# C: Additional Figures

**Abbreviation table for all figures**

| Abbreviation | Definition |
|---|---|
| BEV | Bevacizumab |
| BRE | Breast cancer |
| CER | Cervical cancer |
| CHM | Chemotherapy |
| CI | Confidence interval |
| COL | Colorectal cancer |
| CP | Common parameter |
| Comp | Comparator |
| GLIO | Glioblastoma |
| HMA | Hierarchical meta-analysis |
| HOR | Hormonal therapy |
| HR | Hazard ratio |
| IMM | Immunotherapy |
| IP | Independent parameter |
| NSCLC | Non-small cell lung cancer |
| OFTPP[1] | Ovarian, fallopian tube, and primary peritoneal cancer |
| OS | Overall survival |
| PBO | Placebo |
| PFS | Progression-free survival |
| RAD | Radiotherapy |
| REN | Renal cell carcinoma |
| SE | Standard error |
| TAR | Targeted therapy |

[1] These three cancers were also collectively referred to as 'ovarian cancer'.

**Figure S1.** Timeline plot with start points weighted according to sample size

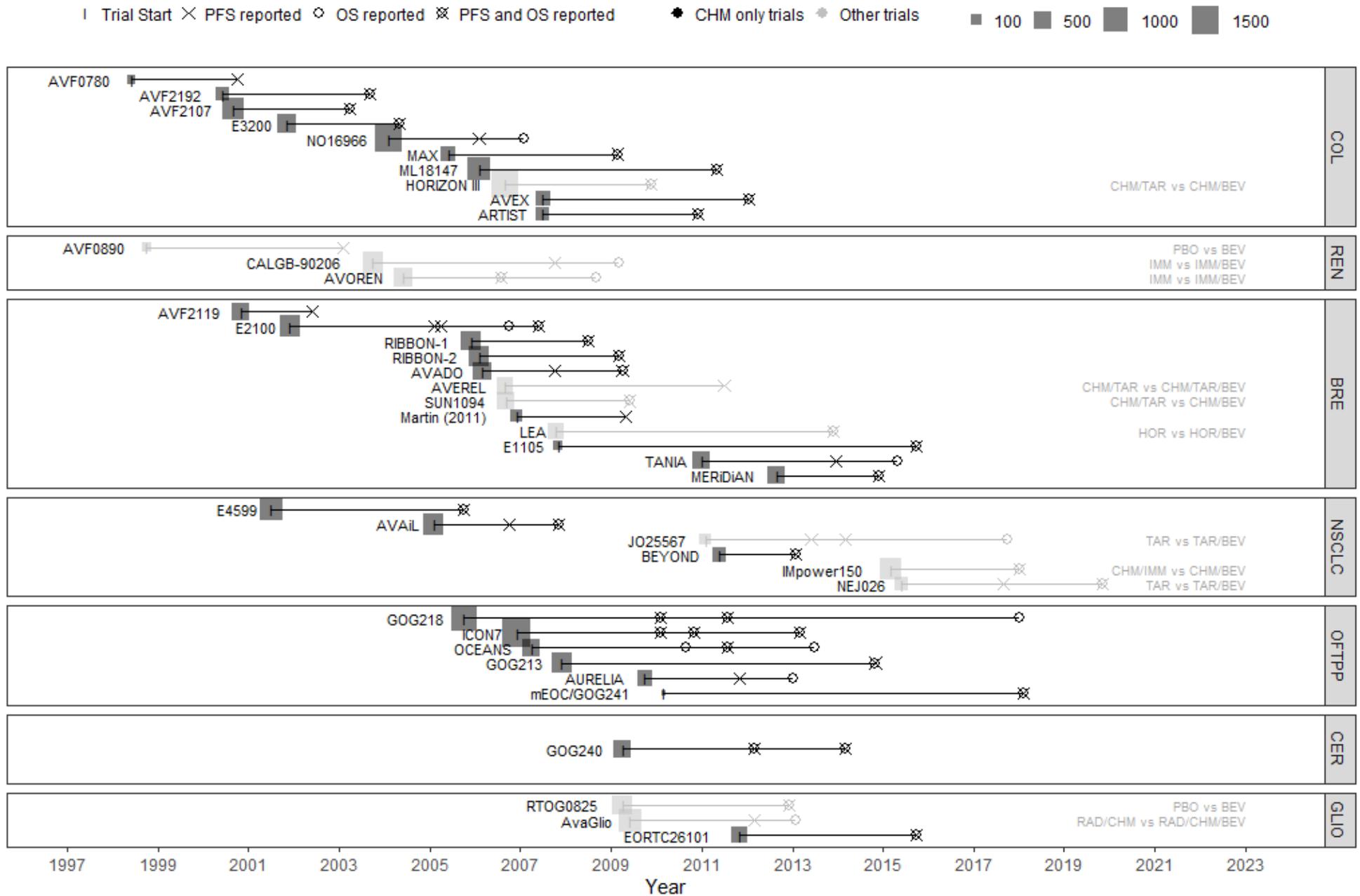

**Figure S2**. Timeline plot showing the maturity of OS evidence

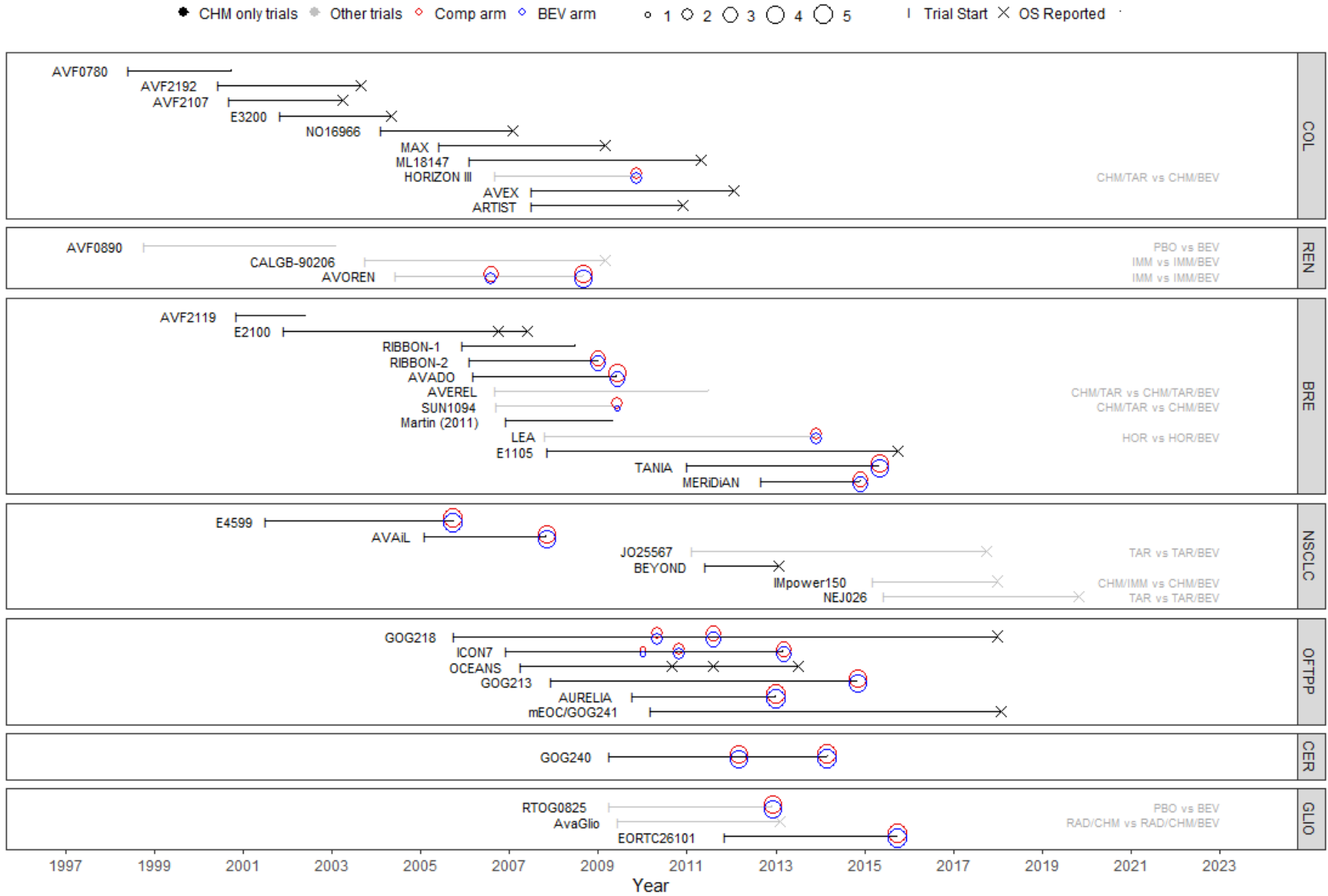

**Key for circle size:** The circles in the legend have the following maturity values (calculated as the proportion of events/total patients) **1**:less than 0.25, **2**: 0.26 to 0.40, **3**: 0.41 to 0.55, **4**: 0.56 to 0.70, **5**: 0.71 and over.

**Figure S3.** Timeline plot showing the maturity of PFS evidence

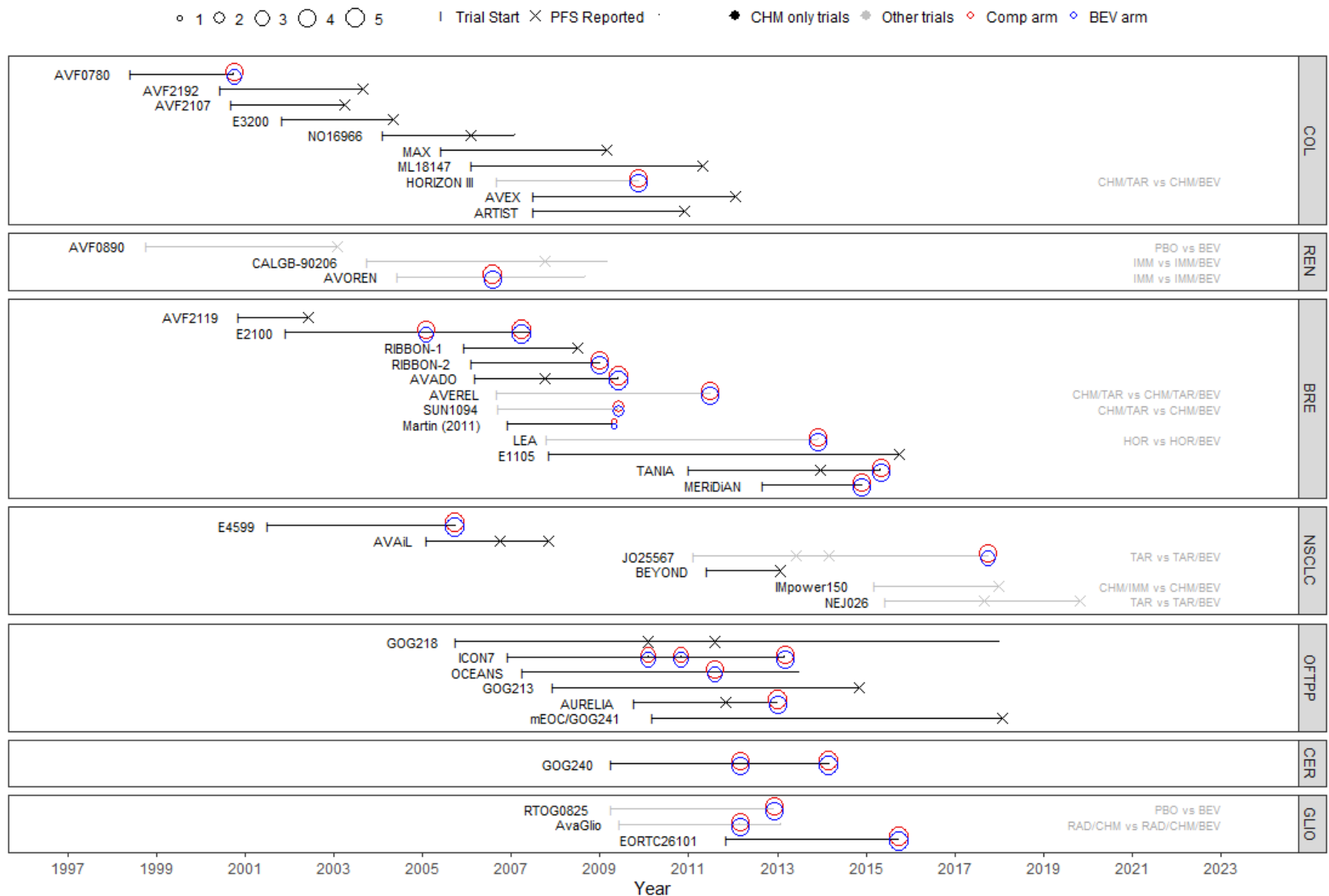

**Key for circle size:** The circles in the legend have the following maturity values (calculated as the proportion of events/total patients) **1**: less than 0.25, **2**: 0.26 to 0.45, **3**: 0.46 to 0.65, **4**: 0.66 to 0.85, **5**: 0.86 and over.

**Figure S4**. Timeline plot showing the uncertainty, measured as the width of CI

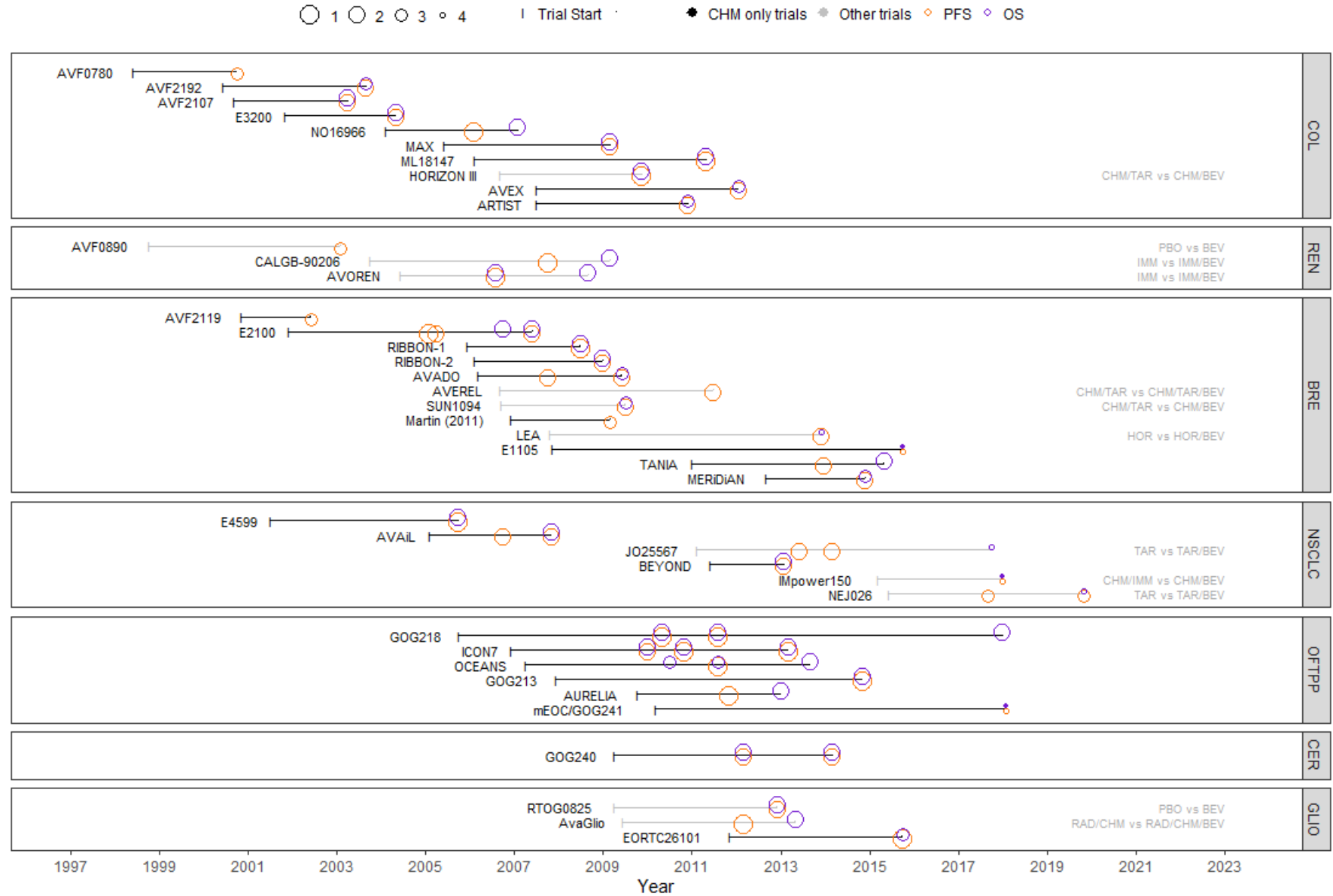

**Key for circle size:** The circles in the legend have the following uncertainty values (calculated as the width of the CI) **1**:less than 0.25, **2**: 0.26 to 0.45, **3**: 0.46 to 0.65, **4**: 0.66 and over. For extreme values of uncertainty (defined as an uncertainty of more than 1.00), the uncertainty is represented by a point in the relevant colour.

**Figure S5**. Timeline plot showing the uncertainty, measured as SE/ln(HR)

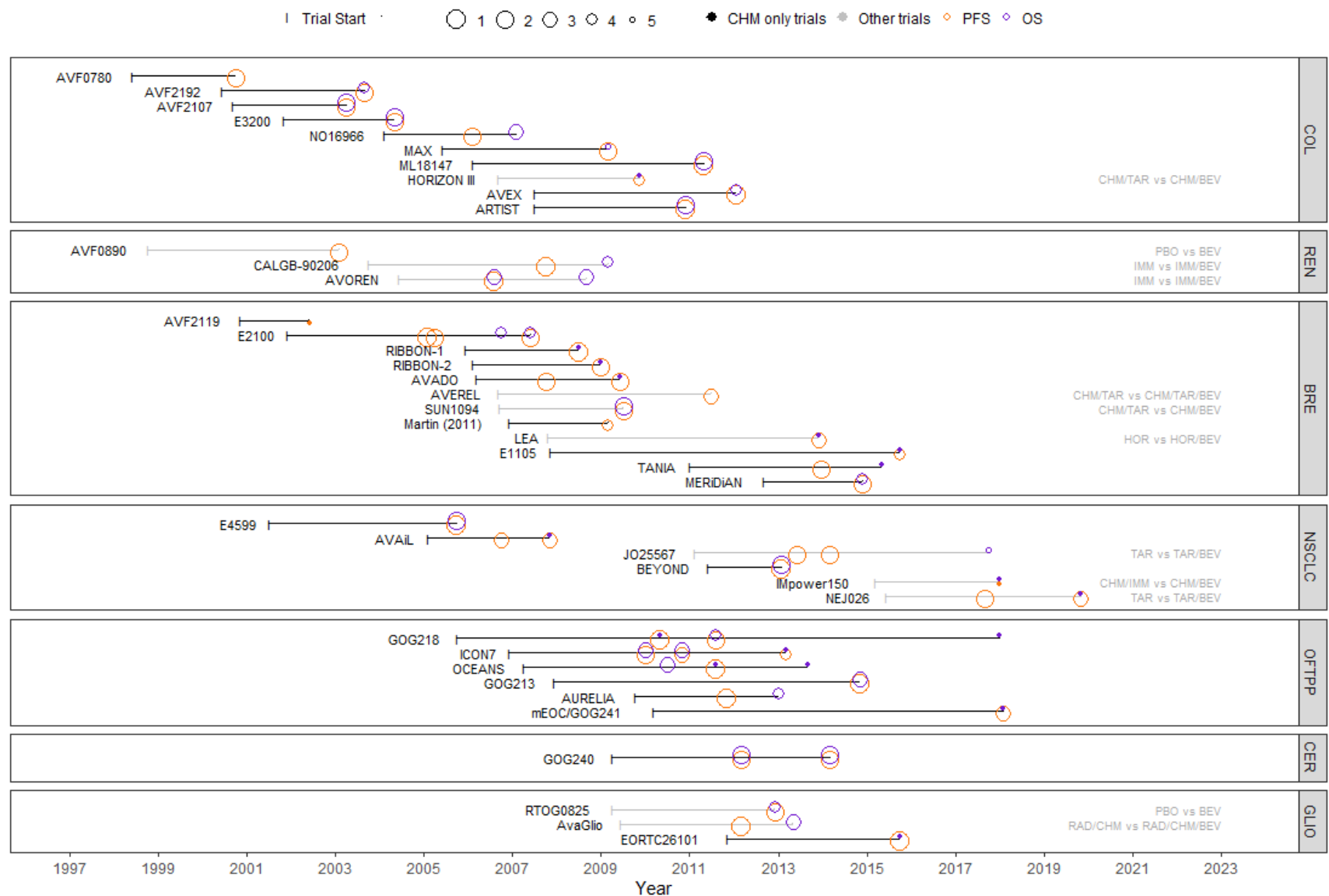

**Key for circle size:** The circles in the legend have the following uncertainty values (calculated as the width of the CI) **1**:less than 0.25, **2**: 0.26 to 0.45, **3**: 0.46 to 0.65, **4**: 0.65 to 1.00, **5**: 1.00 and over. For extreme values of uncertainty (defined as an uncertainty of more than 1.50), the uncertainty is represented by a point in the relevant colour.

**Figure S6**. Ridgeline plots of studies ranked by largest OS

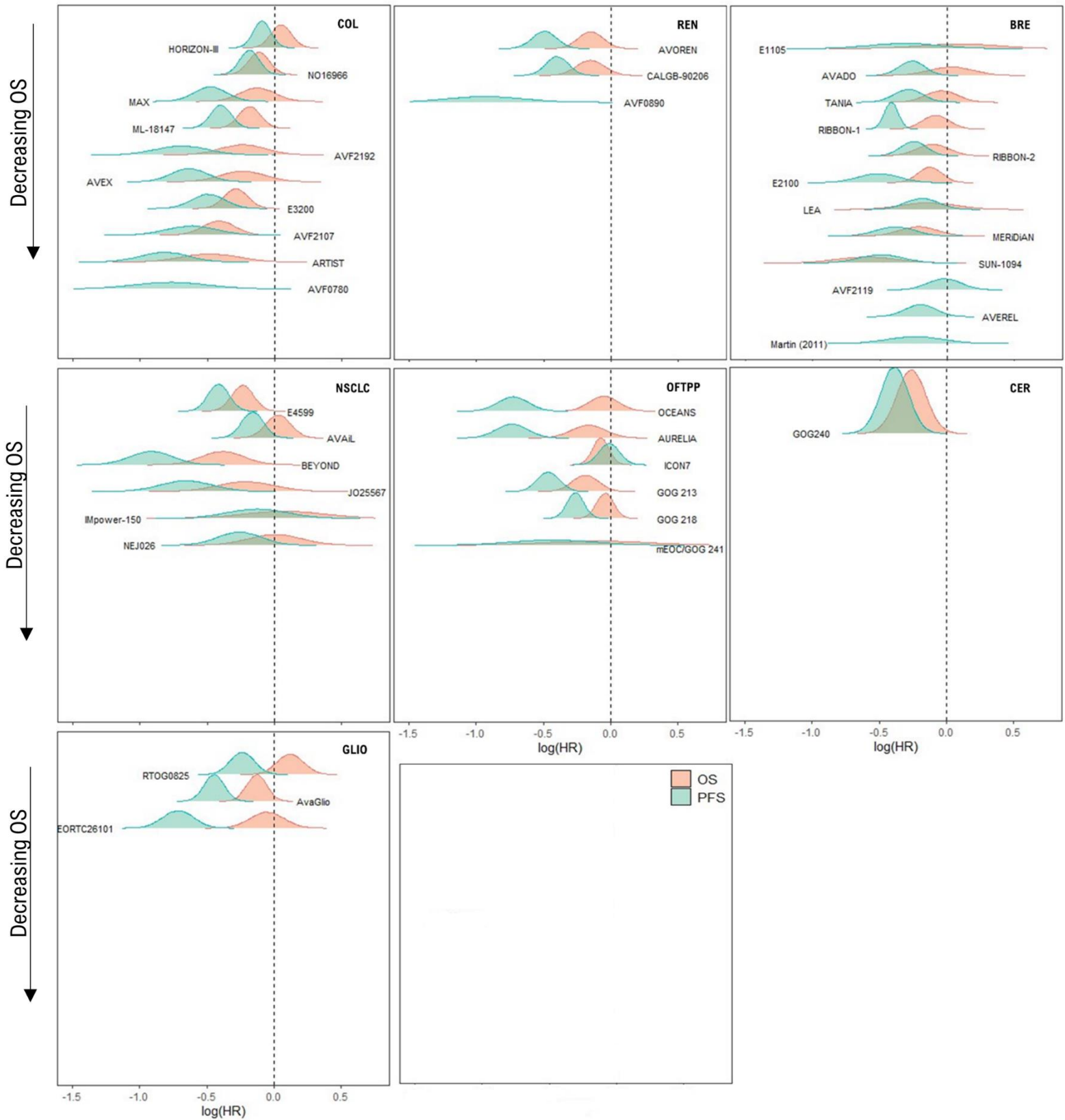

**Figure S7**. Cumulative ridgeline plots comparing meta-analysis models for overall survival

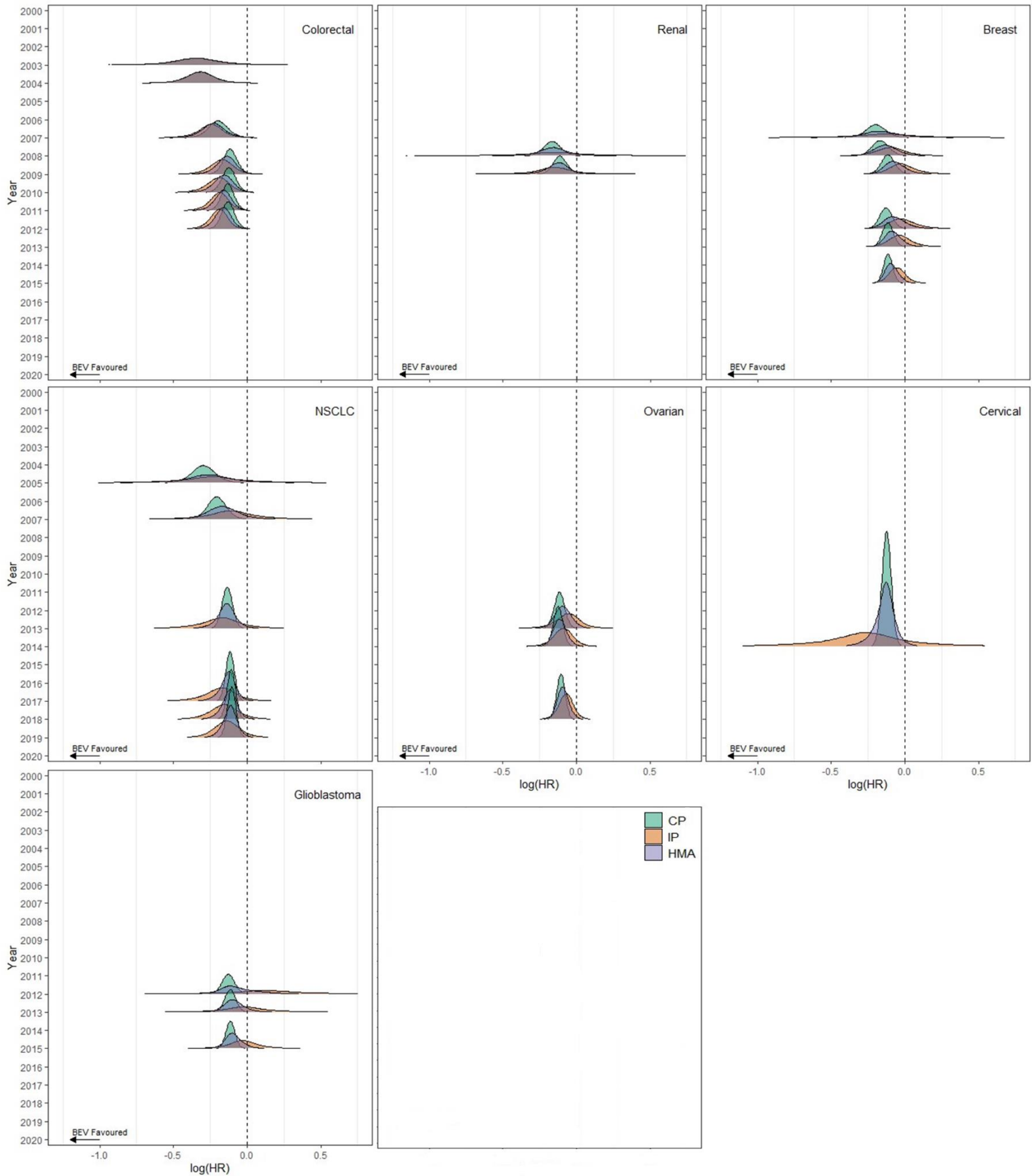

**Figure S8**. Cumulative ridgeline plots comparing meta-analysis models for progression-free survival

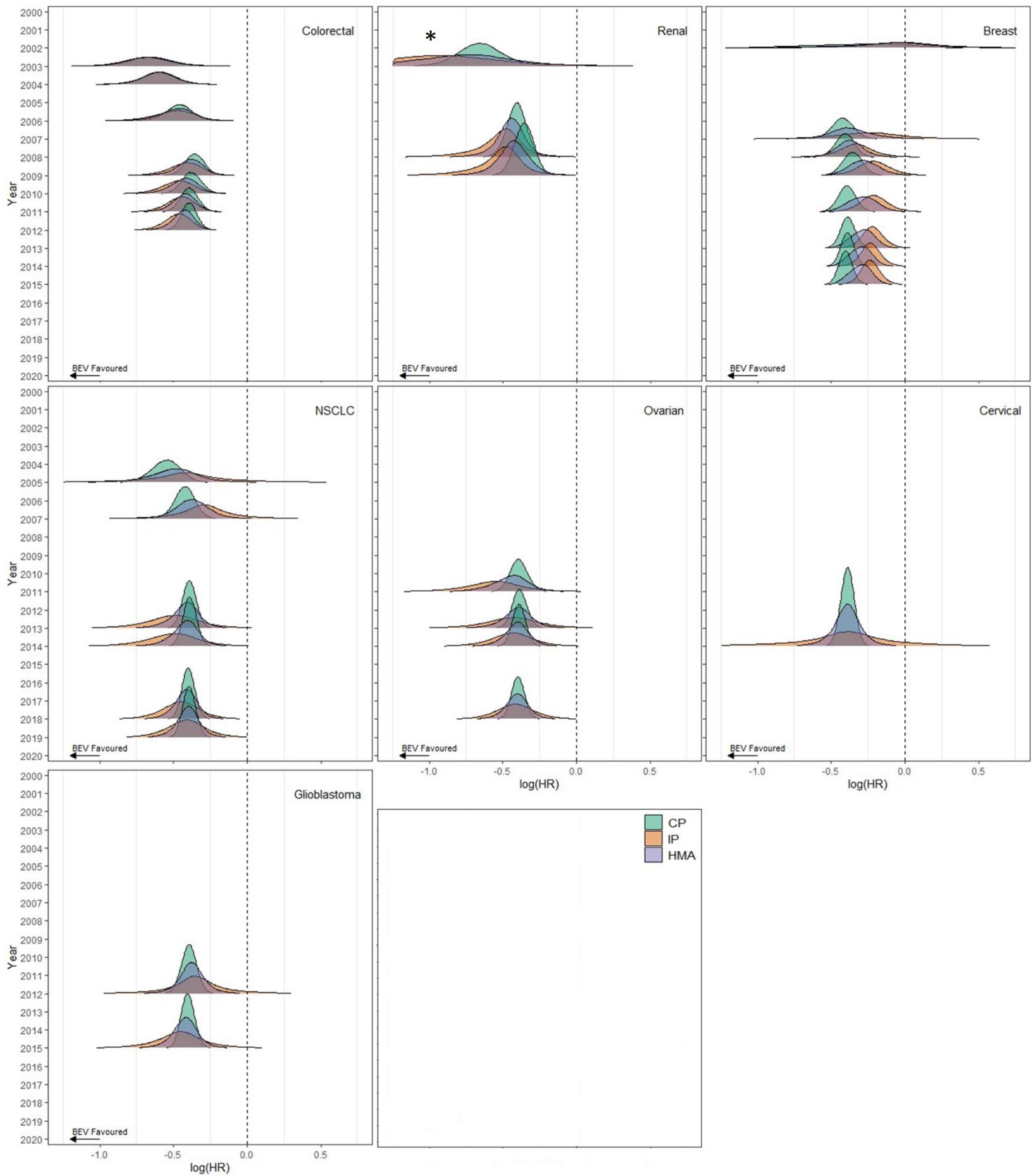

*Density curves are cut-off for display purposes.

# D: References